\documentclass[a4paper,fleqn,usenatbib]{mnras}
\pdfoutput=1
\usepackage{newtxtext,newtxmath}

\usepackage[T1]{fontenc}
\usepackage{ae,aecompl}

\usepackage{graphicx}	
\usepackage{amsmath}	
\usepackage{amssymb}	
\usepackage{natbib}

\newcommand{\kms}{\,km\,s$^{-1}$}    
\newcommand{\Msun}{$M_{\sun}$\,}      
\newcommand{\Lsun}{$L_{\sun}$\,}      
\newcommand{\Rsun}{$R_{\sun}$\,}      
\newcommand{\Msuny}{$M_{\sun}$\,yr$^{-1}$}      
\newcommand{\Teff}{$T_\text{eff}$~}      
\newcommand{\vsini}{$v \sin i$}      

\title[Accreting T Tauri stars RY Tau and SU Aur]{Dynamics of wind and the dusty environments in the accreting T Tauri stars RY Tau and SU Aur \\ }

\author[P. P. Petrov et al.]{P. P. Petrov $^{1}$\thanks{E-mail: petrov@craocrimea.ru},
K. N. Grankin$^{1}$,
J. F. Gameiro$^{2,3}$,
S. A. Artemenko$^{1}$,
E. V. Babina$^{1}$,
\newauthor R. M. G. de Albuquerque$^{2,3,4}$,
A. A. Djupvik$^{5}$,
G. F. Gahm$^{6}$,
V. I. Shenavrin$^{7}$,
\newauthor T. R. Irsmambetova$^{7}$,
M. Fernandez$^{8}$,
D. E. Mkrtichian$^{9}$,
and S. Yu. Gorda$^{10}$
\\
$^{1}$Crimean Astrophysical Observatory of Russian Academy of Sciences, p/o Nauchny, 298409, Republic of Crimea \\
$^{2}$Instituto de Astrof\'isica e Ci\^encias do Espa\c{c}o, Universidade do Porto, CAUP, Rua das Estrelas, PT4150-762 Porto, Portugal \\
$^{3}$Departamento de F\'isica e Astronomia, Faculdade de Ci\^encias, Universidade do Porto, Rua do Campo Alegre 687, PT4169-007 Porto, Portugal \\
$^{4}$Laboratoire Univers et Th\'eories, Observatoire de Paris,  UMR 8102 du CNRS, Universit\'e Paris Diderot, F-92190 Meudon, France \\
$^{5}$Nordic Optical Telescope, Rambla Jos\'e Ana Fern\'andez P\'erez, 7, 38711 Bre\~na Baja, Spain \\
$^{6}$Department of Astronomy, AlbaNova University Center, Stockholm University, Sweden \\
$^{7}$Sternberg Astronomical Institute, M. V. Lomonosov Moscow State University, Moscow, Russia \\
$^{8}$Institute of Astrophysics of Andalusia-CSIC, Glorieta de la Astronomía, 3, 18008 Granada, Spain\\
$^{9}$National Astronomical Research Institue of Thailand, 260 Moo 4, T. Donkaew, A. Maerim, Chiangmai, 50180 Thailand\\
$^{10}$Ural Federal University, 51, Lenin av., Ekaterinburg, Russia, 620000}

\date{Accepted XXX. Received YYY; in original form ZZZ}

\pubyear{2018}

\begin{document}
\label{firstpage}
\pagerange{\pageref{firstpage}--\pageref{lastpage}}
\maketitle

\begin{abstract}
Classical T Tauri stars with ages of less than 10 Myr possess accretion discs. Magnetohydrodynamic processes at the boundary between the disc and the stellar magnetosphere control the accretion and ejections gas flows. We carried out a long series of simultaneous spectroscopic and photometric observations of the classical T Tauri stars RY Tau and SU Aur with the aim to quantify the accretion and outflow dynamics at time scales from days to years. It is shown that dust in the disc wind is the main source of photometric variability of these stars. In RY Tau we observed a new effect: during events of enhanced outflow the circumstellar extinction gets lower. 
The characteristic time of changes in outflow velocity and stellar brightness indicates that the obscuring dust is near the star. 
The outflow activity in both stars is changing on a time scale of years. Periods of quiescence in H$\alpha$ profile variability were observed during 2015-2016 season in RY Tau  and during 2016-2017 season in SU Aur.  We interpret these findings in the framework of the magnetospheric accretion model, and discuss how the global stellar magnetic field may influence the long-term variations of the outflow activity. 

\end{abstract}



\section{Introduction}


Young stars possess accretion discs at the beginning of their evolution. For solar mass stars the lifetimes of these discs are a few million years. Eventually the disc dissipates and the mass accretion ceases. Classical T Tauri stars (cTTS) are young low mass stars (M < 2-3 \Msun) with accretion discs. Their characteristic  emission line spectrum, irregular light variability, and non-stationary outflows are consequences of the interaction of the accreting gas with the stellar magnetic field. When accretion ceases, the outflows disappear and the emission line spectra become weak. These are the weak-line T Tauri stars (wTTS). This phase is much longer, 10-100 Myr, until the star eventually reaches the main sequence. 

 The current  view of cTTS and their environments  is based on the magnetospheric accretion model, initially designed for neutron stars \citep{ghosh1979} and later applied for cTTS \citep{camenzind1990,koenigl1991}. According to the models, the accretion disc of a cTTS is truncated by a stellar magnetic field at a distance  of a few stellar radii. The magnetic field penetrates the inner part of the disc and thus the gas of the disc can flow down to the stellar surface along the field lines. A shock is formed at the base of the accretion channel. X-ray and UV  radiation of the accretion shock ionize the infalling gas,  thus giving  rise to the characteristic emission line spectrum of a cTTS. The mass accretion rates in cTTS are in the range of $10^{-10}$ to $10^{-7}$  \Msuny.  The inner accretion disc radii in cTTS, measured by near-infrared interferometry, are typically within 0.1-0.3 AU \citep{millan-gabet2007}. 
Observed properties and models of cTTS can be found in reviews by e.g. \citet{petrov2003}, \citet{bouvier2007}, and \citet{hartmann2016}.

 Besides the accretion flows, cTTS are also characterized by powerful outflows. Large scale ordered magnetic fields are thought to play a key role in forming these gas flows. Different types of  winds have been considered, including a stellar wind flowing along the open magnetic field lines in the polar regions of the star \citep{cranmer2009}, and a disc wind, starting from the extended surface of the accretion disc and accelerated by the magnetic centrifuge of the rotating disc \citep{blandford1982,pudritz1986,matt2005}. Other possible types of winds originate from the inner region of the disc, at the boundary between  the disc and the magnetosphere,  either as a so called ``X-wind'' \citep{shu1994} or a "conical wind" \citep{romanova2009}. The region where the magnetic flux connects the disc with the star is very unstable, where magnetospheric ejections of plasma  can take place \citep{goodson1997,zanni2013}. Magneto-hydrodynamic (MHD) simulations of the accretion and outflow processes in cTTS have been performed by several groups  (see reviews by \citet{bouvier2007}, \citet{romanova2015} and references therein).  The gas flows can be traced by analysis of emission line profiles in spectra of cTTS.  Specific profiles of Hydrogen and Helium lines were calculated for different wind models (e.g., \citealt{kurosawa2011,grinin2012}).  The disc wind also contributes to the irregular light variability of cTTS. Beyond the dust sublimation radius, the disc wind is dusty, which causes circumstellar extinction of cTTS.

The  processes of accretion and the accretion-driven winds are non-stationary. Dynamics of the gas flows depends of the conditions at the boundary between the disc and stellar magnetosphere. The stellar magnetosphere may be not axisymmetric, which causes rotational modulation of the observed flows. The rotational modulations in the emission lines were observed in some cTTS, e.g. in RW Aur A \citep{petrov2003} and also simulated in the MHD models (e.g \citealt{romanova2007}).

Long term variations in  the outflow activity may be controlled by the variable  mass accretion rate and/or  gradual change in the global stellar magnetic field.   Variations in the magnetic fields of cTTS were reported by  \citet{donati2011,donati2012}.
 
In this paper we present  results of our spectroscopic monitoring of two cTTS, SU Aur and RY Tau, during several years. Besides  long-term changes in the wind activity, we were interested to know whether short-term wind dynamics is reflected in the irregular variations of the circumstellar extinction. For that reason, the spectroscopy was supported by simultaneous photometry of the stars. Preliminary results of the first two years of our monitoring of RY Tau were previously published in \citet{babina2016}.

 The paper is organized as follows. In Sect. 2 we start with review of basic  data for SU Aur and RY Tau and proceed with a description of our observations in Sect. 3.
 The results  obtained  are given in Sect. 4 and discussed in Sect. 5. Finally, the conclusions are listed in Sect. 6.

\section{Basic data for SU Aur and RY Tau}

In this Section we compare  observed characteristics of RY Tau and SU Aur to outline their similarities and differences. Both stars belong to intermediate mass classical T Tauri stars. They are located in the Taurus-Auriga star forming region at a distance of about 140 pc \citep{elias1978,loinard2007}. 
The basic parameters of the stars according to \citet{calvet2004} are \Teff = 5945 $\pm$ 142.5 K for both stars and
stellar luminosities L = 9.6 $\pm$ 1.5 \Lsun in RY Tau and  7.8 $\pm$ 1.3 \Lsun  in SU Aur. 

 The parallaxes measured by GAIA\footnote{\url{http://gaia.ari.uni-heidelberg.de/singlesource.html}} give the following distances: $142.4 \pm 12$ pc for SU Aur and $176.6 \pm 27$ pc for RY Tau , i.e. the distance to  RY Tau could be somewhat larger than usually assumed. 
However, one should take into account the most accurate measurements of parallaxes of several TTS in the Taurus complex obtained in the multi-epoch VLBA array observations \citep{loinard2007}. They found a mean distance to the star-forming region of about 140 pc with a depth of around 20 pc. Hence, the upper limit of the distances measured by VLBA (150 pc) coincides with the lower limit of the distance to RY Tau (149.6 pc) obtained by GAIA. Therefore, we adopt a distance to RY Tau of 150$\pm$10 pc.

\subsection{Light variability}

Both objects have long photometric records.  RY Tau has shown irregular variability within $V=9.5 -11.5$, with noticeable brightening in 1983-1984 and 1996-1997 \citep{herbst1984,herbst1994,zajtseva1996}. The most extended series of photometric observations of RY Tau from 1965 to 2000 was analysed by \citet{zajtseva2010}.  Quasi-periodic variations of brightness, probably associated with eclipses by dust cloud in the circumstellar disc were revealed. No periodicity related to rotation of the star itself was detected.  

SU Aur is an irregular variable, most of time at $V=9.0-9.5$ with random drops down to $V=10 - 11$ \citep{herbst1999,dewarf2003}. Possible periods of 1.55 and 2.73 days were reported by \citet{herbst1987}, but not confirmed later \citep{herbst1988}. \citet{bouvier1988,bouvier1993} reported a possible period of 2.78 days, which is close to the rotational period derived from spectral lines variations (see below).  MOST photometry \citep{cody2013} showed small amplitude ($\sim$0.1 mag) brightness oscillations with period of  $\sim$2.7 days  over 20 days of observation.

\subsection{Emission line spectra}

The emission line spectra of RY Tau and SU Aur are not as strong as in late-type cTTS, because of the luminous background of the G-type photosphere. In the optical spectrum of SU Aur only H$\alpha$ is in emission.  In RY Tau the emission spectrum includes H$\alpha$, H$\beta$, the \ion{Na}{I} doublet, \ion{He}{I} 5876 \AA, \ion{Ca}{II} doublet and the NIR \ion{Ca}{II} triplet (e.g.\citealt{hamann1992, alencar2000}).  Forbidden emission lines of [\ion{O}{I}], [\ion{N}{II}], [\ion{S}{II}] and [\ion{Fe}{II}] were observed in the spectrum of RY Tau \citep{cabrit1990,akitaya2009}. No forbidden lines have been reported for SU Aur.

The photospheric spectrum of a cTTS is often veiled by additional radiation from hot surface areas at the base of accretion columns. The effect is stronger in a late type cTTS where the brightness contrast of the hot area is larger. 
The veiling of the photospheric spectrum of RY Tau in the visual range is very low or absent \citep{basri1991,hartigan1995,petrov1999,chou2013}. No veiling in the visible photospheric spectrum of SU Aur has been reported. 
 Excess continuous radiation is present in the  far UV spectrum of both stars,  
larger in RY Tau than in SU Aur. The mass accretion rates, estimated from the accretion luminosities in UV, are 6.4 -- 9.1 $\pm$ 4.9 for RY Tau, and 0.5 -- 0.6 $\pm$ 0.4 for SU Aur, in units of 10$^{-8}$ \Msuny \citep{calvet2004}. 

Near and Far-UV spectra of  TTS, including RY Tau and SU Aur, were analysed by \citet{ardila2002,gomez2007,gomez2012}.   
A catalogue of selected atomic and molecular  line fluxes, observed in the Far IR, is presented in \citet{alonso-martinez2017}.


\subsection{Rotation and X-rays}

Both stars are rapid rotators. In SU Aur, the \vsini \, is within = 60 - 66 \kms \citep{nguyen2012,krull1996,petrov1996}. In RY Tau, the \vsini\ 
$ \sim$ 50 \kms \citep{petrov1999}. The rotation period of SU Aur is within 2.7 - 3.0 days, as determined from periodical modulations of the blue- and red-shifted absorption components in the Balmer line profiles \citep{giampapa1993,johns1995,petrov1996} and strength of \ion{He}{I} emission lines \citep{unruh2004}.  In RY Tau no rotation period was detected, neither from the extended photometric series or from the emission lines variations.  Both RY Tau and SU Aur are X-ray sources. RY Tau is a strong, flaring X-ray source, indicating radiation from  a hot plasma at T $\sim$ 50 MK. The flaring component is undoubtedly of magnetic origin \citep{skinner2016}. The quiescent X-ray emission from SU Aur is dominated by a 20 - 40 MK plasma, while an extremely high temperature plasma component (at least 60 MK)  was observed in a flare \citep{smith:favata2005}.

\subsection{Discs and jets}

Both stars possess accretion discs, as indicated by their spectral energy distribution (SED) and interferometry in the infrared (IR). The SED of RY Tau and SU Aur from SPITZER mid-IR observations \citep{furlan2011} represents radiation of warm layers from an inner disc within 10 AU. 

A number of accretion disc models of RY Tau and SU Aur has been presented to reproduce the IR  observations, e.g. \citet{muzerolle2003}, \citet{akeson2005}, \citet{schegerer2008}, \citet{isella2010},  and \citet{guilloteau2011}.  The inner disc radius in different models drops within 0.3 - 0.5 AU, and  inclinations of the disc axis to the line of sight appear within 55 - 75 \degr.

Imaging polarimetry of SU Aur has shown that the accretion disc extends out to 500 AU with an inclination  of $\sim$ 50$\degr$ \citep{jeffers2014};  the disc morphology with tidal tails was reconstructed by \citet{deleon2015}. Imaging polarimetry of RY Tau  \citep{takami2013} showed that the scattered light in the near-IR is associated with an optically thin and geometrically thick layer above the disc surface. 
The changes in the linear polarization across the H$\alpha$ line in RY Tau and SU Aur are consistent with the presence of a compact source of line emission 
that is scattered off a rotating inner accretion disc \citep{vink2005}. 

Extended bipolar jets of RY Tau, with young dynamical ages of the inner knots, were detected in H$\alpha$ light \citep{onge2008}
and  mapped in [\ion{O}{I}] 6300 \AA\ \citep{agra-amboage2009}. Spatially resolved \ion{C}{IV} emission from the blue-shifted jet of RY Tau  has been detected by  \citet{skinner2018}. 
  
In summary, RY Tau and SU Aur are rather similar  with regard to stellar parameters, but with different accretion rates and circumstellar environments.  The high inclinations of both stars implies that the line of sight intersects the disc winds.  This provides an opportunity to search for dynamics of the circumstellar gas flows  through variability in  spectral line profiles.

\subsection{Spectral time series}

A typical characteristics of the gas flows in cTTS is their non-stability on  time scales of a day and longer. In some cases, spectral monitoring can reveal  modulation of a line profile with a period of the stellar rotation, which may be due to axial asymmetry of the gas flows.

Several extended time series  of high-resolution spectroscopic observations have been obtained for SU Aur \citep{giampapa1993,johns1995,petrov1996,oliveira2000,unruh2004}. The Balmer line profiles show the largest variability in the blue-shifted and red-shifted depressions of the broad emission profiles, formed in the outflow (wind) and inflow (accretion) gas  streams. Variations in the blue wing of H$\alpha$ and H$\beta$ revealed rotational period about 3.0 days \citep{giampapa1993,johns1995,petrov1996}. A shorter period of  2.6 to 2.8 days was found in variations of the \ion{He}{I} D3 line \citep{unruh2000,unruh2004}. Time-series involving Pa$\beta$ spectroscopy of SU Aur over three consecutive nights showed relatively strong variability in the red wing within a radial velocity range of 100 - 420 \kms, and less variability in the blue wing. A model with an inclined dipole magnetosphere reproduced the observed line variability \citep{kurosawa2005}.  Sometimes a strong depression of the blue wing of H$\alpha$ appeared, indicating  enhanced sporadic  mass ejections \citep{petrov1996}.   

Time variability of the emission lines in RY Tau were studied by \citet{zajtseva1985}, \citet{petrov1990}, \citet{johns1995}, \citet{smith1999}, \citet{mendigutia2011}, and \citet{chou2013}. The broad emission profiles of the H$\alpha$ line is cut by a deep blue-shifted depression. Both the blue and red peaks  of the line are variable on a time scale of a few days. Balmer H$\alpha$ line monitoring of RY Tau in 37 almost fully contiguous nights \citep{johns1995} did not reveal any periodicity similar to SU Aur.  Although both RY Tau and SU Aur are fast rotators with similar stellar parameters,  axial rotation of RY Tau  is not reflected as variability in the Balmer line profiles.  \citet{ismailov2015} reported possible period of 23 days in variability of the \ion{Mg}{II} 2800 \AA\ emission line intensity in IUE spectra of RY Tau.  This is close to the  periods found from photometric series \citep{bouvier1993,gahm1993}.  This long  period is certainly not related to the stellar rotation.

\section{Observations}

Our spectral and photometric observations were carried out during five seasons of 2013-2018. We started with observations of RY Tau at the Crimean Astrophysical Observatory (CrAO) in the 2013-2014 and 2014-2015 seasons. Then, a multi-site monitoring of RY Tau and SU Aur was arranged  in 2015-2016 season. Later we proceeded with observations of RY Tau and SU Aur, mostly at CrAO. In all the seasons, photometry of our targets were performed at three telescopes, located in Crimea.
\\

\subsection{Spectroscopy}

The following  five instruments were used to obtain series of spectral observations of RY Tau and SU Aur.  

(1) 2.6-m Shajn reflector of the Crimean Astrophysical Observatory (CrAO) with echelle spectrograph. Spectral resolution R=27000 with entrance slit 2$\arcsec$.

(2) 2.5-m Nordic Optical Telescope (NOT) with  ALFOSC grism spectrograph, grism set \#17, registered spectral region 6330-6870\AA\ ,  R=10000 with entrance slit 0.5$\arcsec$.

(3) 2.2-m telescope at Centro Astron\'omico Hispano-Alem\'an (CAHA) with CAFE echelle spectrograph, R=58000  with entrance slit 1.2$\arcsec$.

(4) 2.4-m Thai National Telescope (TNT) at Thai National Observatory (TNO) with Medium Resolution Spectrograph (MRES), R=19000 with entrance fiber 2 $\arcsec$.

(5) 1.2-m telescope of Kourovka Astronomical Observatory of the Ural Federal University (UrFU), with fiber-fed echelle spectrograph, R=15000, entrance fiber 5 $\arcsec$ \citep{panchuk2011}.

Further details on the spectrographs can be found at the corresponding web-sites of the observatories. All spectra were reduced and wavelength calibrated using standard procedures and IRAF tools. In the echelle spectra we utilized only those spectral orders which cover regions of the H$\alpha$ and Na D lines.  Signal-to-noise ratio per resolution element at the continuum level in all the spectra was over 100. Total number of nights of spectral observations: 127  for RY Tau and 96  for SU Aur. Log of spectral observations is given in Table 1A and Table 2A.

\subsection{Photometry}\label{photometry}

Optical photometry of RY Tau and SU Aur in the Johnson BVRI photometric system was carried out with two instruments: 1.25-m telescope (AZT-11) of the Crimean Astrophysical Observatory (CrAO), and 0.6-m telescope (Zeiss-600) at the Crimean Astronomical Station (CAS) of Moscow State University. At the 1.25-m telescope, photometer with FLI PL23042 CCD camera was used for a routine differential photometry, while in cases of perfectly clear sky a photo-counting photometer was used for absolute photometry. The standard error  is about 0.02 mag in all bands. At the 0.6 m telescope, photometer with CCD cameras Apogee Aspen and PL4022 was used.

Near Infrared (NIR) photometry was carried out at the 1.25-m telescope of  CAS.  InSb-photometer with a standard JHKLM system was used. Technical parameters of the photometer, methods of observations and calculations of magnitudes were described in details by \citet{shenavrin2011}. Stars BS1203 and BS1791  were used as standards for RY Tau and SU Aur correspondingly. JKL magnitudes of the standards were taken from the catalogue of \citet{johnson1966}, and HM magnitudes were calculated from relations given in \citet{koornneef1983}. The standard error of the measured magnitudes  is about 0.02 in JHKL bands, and about 0.05 in M band.  
In addition, we used also  photometric data  from  the American Association of Variable Star Observers (AAVSO) \citep{kafka2017}.

\begin{figure}
	\includegraphics[trim={1cm 1cm 0 0},clip,width=1.2\columnwidth]{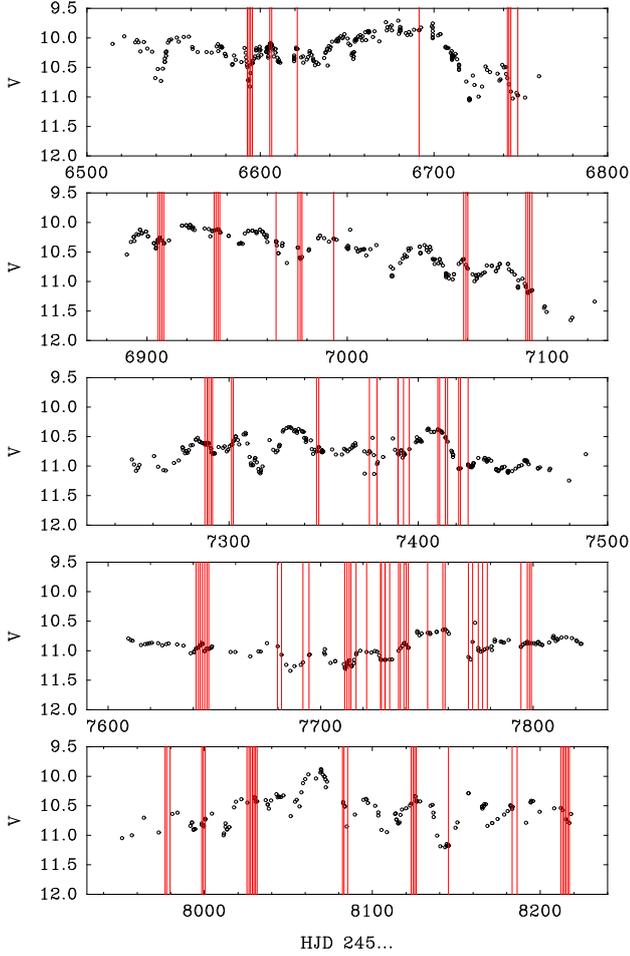}
	\caption{Light curves of RY Tau in five seasons, from 2013-2014 (upper panel) to 2017-2018 (lower panel). Vertical lines mark the moments of spectral observations. }
	\label{fig:RY_13_18}
\end{figure}

\begin{figure}
	\includegraphics[trim={1cm 11cm 0 0},clip,width=1.25\columnwidth]{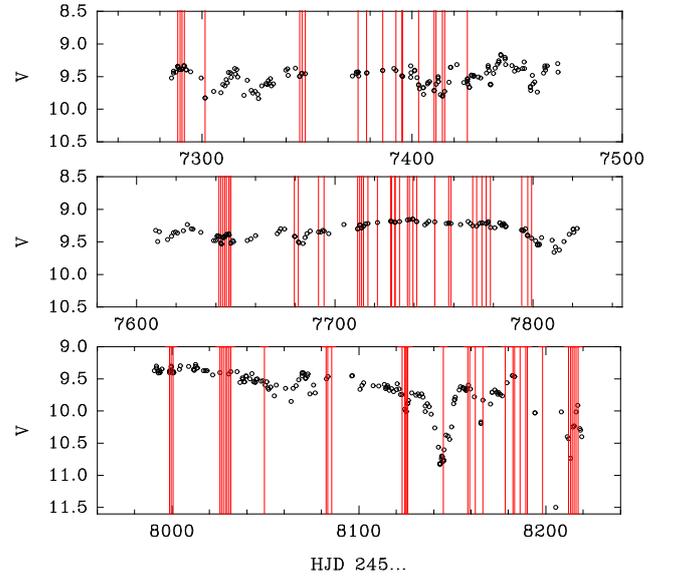}
	\caption{Light curves of SU Aur in three seasons, from 2015-2016 (upper panel) to 2017-2018 (lower panel). Vertical lines mark the moments of spectral observations. }
	\label{fig:SU_15_18}
\end{figure}

The results of the photometric observations in V-band are presented in form of light-curves in Fig.~\ref{fig:RY_13_18} and \ref{fig:SU_15_18}, where moments of spectral observations are marked with vertical bars.  
AAVSO data are added to the figure to make the lightcurves more dense. V-magnitudes in the dates of spectral observations are presented also in Tables~\ref{tab:table02} and \ref{tab:table03} for RY Tau and SU Aur, respectively. In case we had no photometric observation at the date of spectral observation, the V-magnitude was taken either from AAVSO data or from linear interpolation between the nearest dates in the photometric series. The probable error of the interpolated V-mag was roughly estimated as  0.1 mag taking into account  the typical gradients in the light-curves. The results of the NIR photometry for both stars is given in Tables~\ref{tab:table04} and \ref{tab:table05}, where we include also visual photometry, when available at the dates of NIR photometry. Our previous NIR photometry  of RY Tau in 1981-1997 is also included in Table~\ref{tab:table04} .

\section{Results}

\subsection{Photometry}

We start the analysis of the observations with a revision of the basic stellar parameters of RY Tau and SU Aur. From optical photometry we estimate the interstellar extinction $A_\textup{v}$ and calculate the absolute magnitude $M_V$. We adopt  \Teff $=5945 \pm 140$ K \citep{calvet2004}, which corresponds to spectral type G1-G2 IV.  Then, with \Teff and $M_V$ we enter the PMS models by \citet{siess2000} and get the stellar luminosity, mass, radius and age.

\begin{figure}
	\includegraphics[trim={2cm 17.5cm 6cm 3cm},clip,width=\columnwidth]{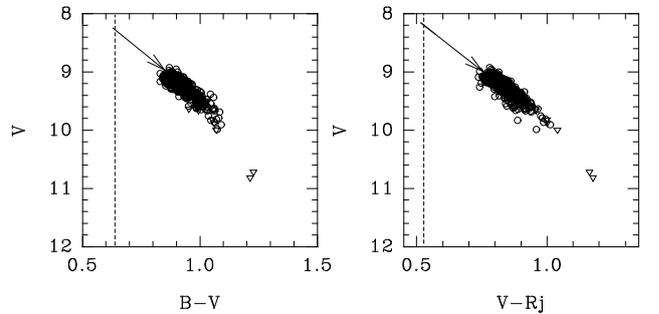}
	\caption{Colour-magnitudes diagrams for SU Aur.  Open circles - Majdanak data of 1983- 2003.  Triangles - CrAO photometry during the miminum of 2017. Arrows indicate the  slope of the interstellar reddening. Vertical dashed lines mark the normal colours of a star with \Teff $=5945$ K.}
	\label{fig:SU_color}
\end{figure}

\begin{figure}
	\includegraphics[trim={2cm 17.5cm 6cm 3cm},clip,width=0.97\columnwidth]{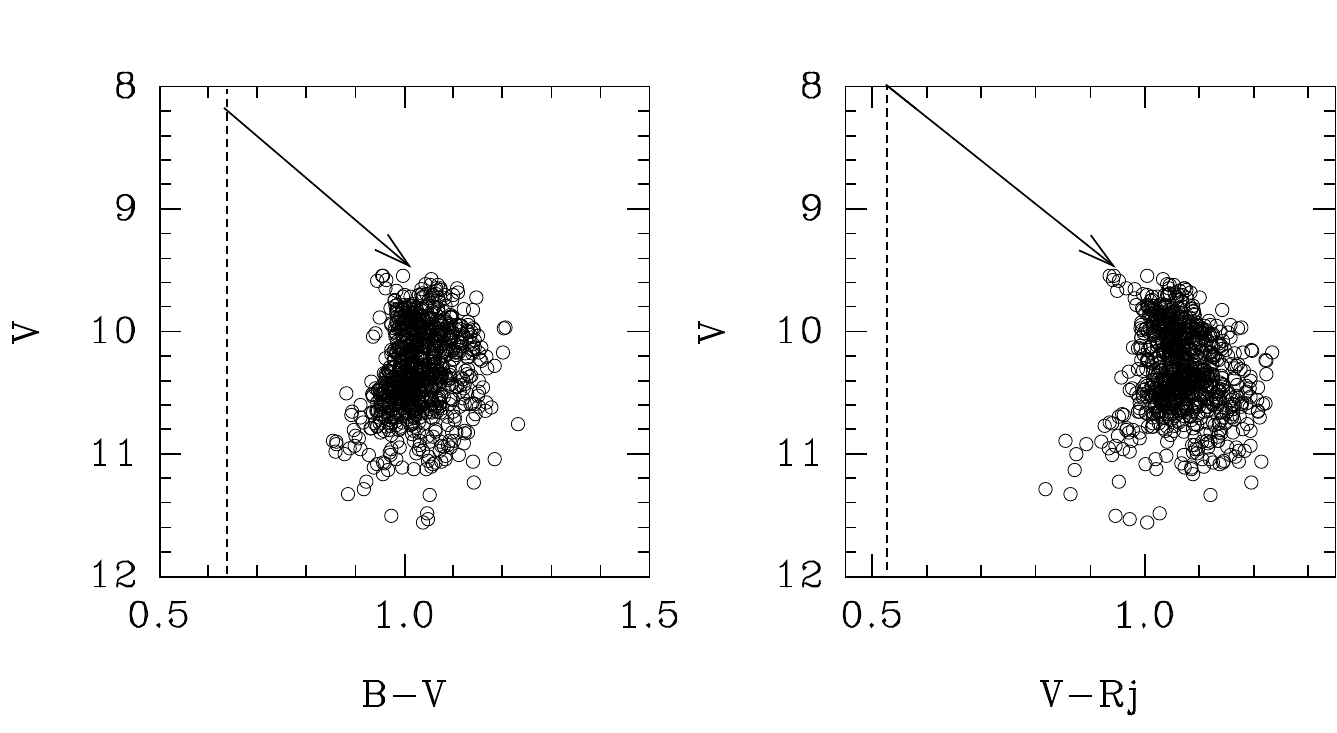}
	\caption{Colour-magnitudes diagrams for RY Tau according to  Majdanak data of 1984 - 2004. Arrows indicate the  slope of the  interstellar reddening. Vertical dashed lines mark the normal colours of a star with \Teff $=5945$ K.}
	\label{fig:RY_color}
\end{figure}

In Fig.~\ref{fig:SU_color} and \ref{fig:RY_color} we plot colour-magnitude diagrams using the most extended and uniform photometric data collected at Majdanak observatory over 20 years \citep{grankin2007}.  
The brightest V magnitudes ($V_\textup{max}$) in the diagrams supposedly represent the normal brightness of the star, not obscured by circumstellar matter. We use ($V-R_j$) colours to estimate the interstellar reddening. A star with \Teff $=5945$ K has  an intrinsic colour of $(V-R_j) = 0.52$ mag \citep{kenyon1995}. The unreddened V magnitudes, corresponding to the intrinsic ($V-R_j$) colour,  are $V=8.20$ mag for SU Aur and $V=8.00$ mag for RY Tau, as indicated by  the cross-section of the reddening lines with the dashed lines. With an error of the adopted \Teff about $142$ K and the scatter of points  in the $V/(V-R_j)$ diagrams, the error of the unreddened V may be within $0.10$ mag.  From these estimates we get the interstellar reddening $A_\textup{v} = 0.80 \pm 0.10$ mag for  SU Aur and $A_\textup{v} = 1.60 \pm 0.10$ mag for RY Tau. These values are consistent with previous estimates of the reddening from optical data (e.g. \citealt{calvet2004,herczeg2014,grankin2017}).

The diagrams of RY Tau show a reversal of colours: as the circumstellar extinction gets large, the colour turns bluer . The colour reversal effect is typical for UX Ori type stars: obscuration of star by the circumstellar dust results in increased contribution of the light scattered on the dust particles. This effect was earlier observed in some  cTTS,
e.g.  RY Lup \citep{gahm1989,gahm1993b}. 
 In RY Tau we do not see the linear part of the $V/(B-V)$ diagram but only the curved track. One may suspect that the star still remains obscured by circumstellar dust even at the brightest state. Therefore, in the case of RY Tau, the estimated $A_V$ may be considered only as an upper limit of interstellar reddening.  In SU Aur this effect is normally absent.  The resulting stellar parameters are given in Table~\ref{tab:table06}.

\begin{table*}
\centering
\caption{Stellar parameters of  RY Tau and SU Aur.}
\label{tab:table06}
\begin{tabular}{cccccccccc} 
\hline
\hline
Star      &  distance  & Vmax   &  Av    &   Mv   &  \Teff $^a$  &  L/\Lsun  &  R/\Rsun  &   M/\Msun   &    Age \\
            &      pc         & mag    &  mag  &   mag &      K             &                    &                     &                       &    Myr    \\
\hline
RY Tau &  $150.0 \pm 10$ & $9.55$ & $1.60$ & $2.11 \pm 0.15$ &  $5945.0\pm142.5 ^a$    &  $13.50 \pm 1.70$ & $3.30 \pm 0.25$ & $2.08 \pm 0.10$ & $4.70^{+1.00}_{-0.80}$ \\
SU Aur &  $142.4 \pm 12$ & $8.95$ & $0.80$ & $2.43 \pm 0.18$ &  $5945.0\pm142.5$   &  $10.09 \pm 1.64$ & $2.85 \pm 0.22$ & $1.88 \pm 0.10$ & $6.60^{+1.04}_{-0.82}$ \\
\hline
\multicolumn{9}{l}{ $^a$  \Teff  is from \citet{calvet2004}}\\

\end{tabular}
\end{table*}

\begin{figure}
	\includegraphics[trim={4cm 13.5cm 4cm 3cm},clip,width=0.75\columnwidth]{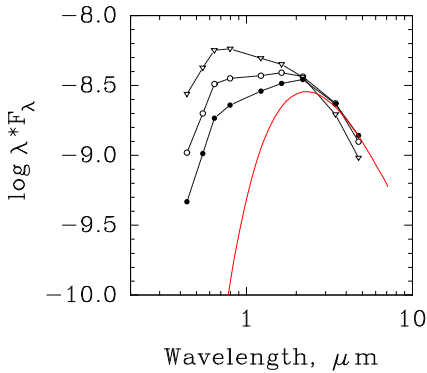}
	\caption{Spectral energy distributions (SED) of SU Aur, corresponding to V=9.18 (triangles), V=10.00 (open circles) and V=10.72 (filled circles).
                         Solid curve: SED of a black body at $T=1600$ K. Flux $F_{\lambda}$  is expressed  in units of erg cm$^{-2}$ s$^{-1}$ \AA$^{-1}$.}
	\label{fig:SU_9}
\end{figure}

\begin{figure}
	\includegraphics[trim={1cm 13.5cm 5cm 3cm},clip,width=0.9\columnwidth]{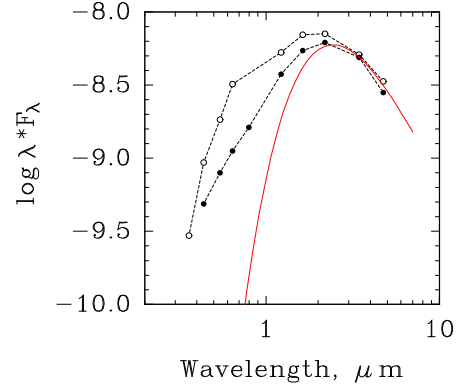}
	\caption{Spectral energy distributions  in  RY Tau, corresponding to V=10 (open circles) and V=11 (filled circles). Solid curve: SED of a black body at T=1500 K. Flux $F_{\lambda}$  is expressed  in units of erg cm$^{-2}$ s$^{-1}$ \AA$^{-1}$.}
	\label{fig:RY_SED}
\end{figure}

The pattern of light variability of RY Tau and SU Aur may be illustrated with the SED  in the optical and NIR regions, covered by our photometry (see Tables~\ref{tab:table04} and \ref{tab:table05}). We selected a few observations at high and low brightness in the V band of each star. The corresponding SEDs are shown in Fig.~\ref{fig:SU_9} and \ref{fig:RY_SED}.  In both stars the SED is a sum of the stellar radiation reddened by the interstellar and circumstellar extinctions and a black-body radiation of the circumstellar dust.  The difference between the two stars is the relative contribution of the stellar and circumstellar radiation.  In RY Tau the optical part of the SED is more depressed by the large circumstellar extinction. At minimal brightness the optical SED is more flat because of the  light scattered by the circumstellar dust.  There is a small but noticable variability in the  NIR part of the SED in both stars.

\subsection{Spectroscopy}

In the analysis of the spectral series we focus on variations of the H$\alpha$ and \ion{Na}{I} D line profiles, which are the strongest indicators of gas flows in the visible region of the spectrum. In addition, some other lines, including photospheric  lines, were involved in the analysis.  Our spectroscopic series cover a major part of the light variations:  from $V=9.8$ mag  to $V=11.2$ mag  in RY Tau and  from $V= 9.3$ mag to $V=10.8$ mag  in SU Aur.  In both stars the photospheric spectrum remains unchanged: the depth of the photosphere lines in the region around 6000\AA\  at  high and low brightness remains the same within $\pm2\%$ of the continuum level.  It means that the observed light variations are not related to any surface phenomenon, like  hot or cool spots, but are mostly due to variable circumstellar extinction.  The stellar radial velocity is $+18.0\pm2.0$ \kms in both RY Tau and SU Aur, consistent with previous measurements  \citep{petrov1996, petrov1999}.

\begin{figure*}
    \includegraphics[trim={1cm 12cm 2cm 2cm},clip,width=1.4\columnwidth]{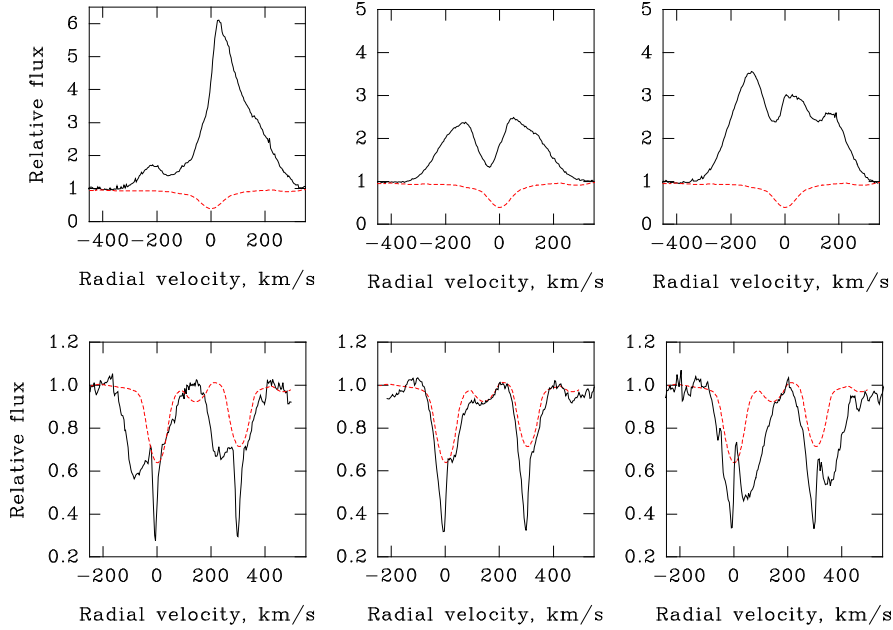}
    \caption{Sample of line profiles in RY Tau. Upper panels: H$\alpha$, lower panels:  \ion{Na}{I} D.  Left column: HJD=2456993.262,  middle: HJD =2457799.442,  right: HJD=2457091.204. Dashed curves:  template  G2 V star,  $v\sin i = 50$ \kms.}
    \label{fig:RY_HA_NA}
\end{figure*}

The most evident result is the large variability in H$\alpha$ and \ion{Na}{I} D profiles on a timescale of a day and longer.  Samples of typical line profiles in RY Tau are shown in Fig.~\ref{fig:RY_HA_NA}. In this and other diagrams of spectral lines we use astrocentric radial velocity scale.  The H$\alpha$ line is in broad emission extending from -300 to +300 \kms,  with a strong ''central'' depression at about -100 \kms, which often extends  further to the blue and in rare cases drops even below continuum. In terms of line profile classification by \citet{reipurth1996}, H$\alpha$   is of  II-B type most of the time. This is characteristic of an outflow, probably a disc wind \citep{kurosawa2011} or conical wind \citep{kurosawa2012}. On the other hand, a red-shifted depression in the H$\alpha$ profile at +100 to +200 \kms is often present. Comparison of the concurrent H$\alpha$  and \ion{Na}{I} D line profiles in Fig.~\ref{fig:RY_HA_NA} clearly shows that the depressions in H$\alpha$ profile correspond to real absorption components  in the \ion{Na}{I} D lines, although at a lower range of velocities. Therefore, the red-shifted absorption in both lines forms in the infalling gas, most probably in the accretion funnels. These red-shifted absorptions, indicating accretion, are much better seen in the Pashen and Brackett series as real absorption below the continuum (e.g. \citealt{folha2001}). In our analysis of the H$\alpha$ profile variability we study only the outflow activity.  

\begin{figure}
    \includegraphics[trim={2cm 11cm 4cm 2cm},clip,width=1.1\columnwidth]{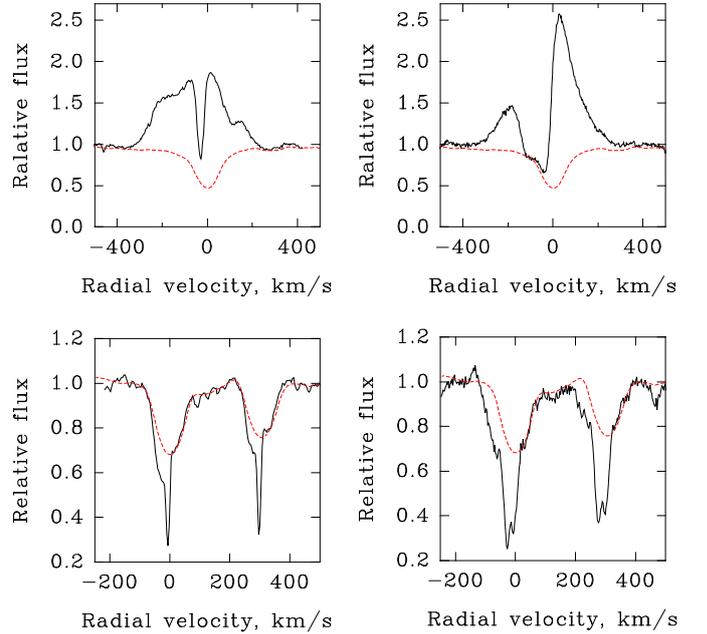}
    \caption{Sample of  line profiles in SU Aur. Upper panels: H$\alpha$, lower panels:  \ion{Na}{I} D.  Left column: HJD=2458030.503,  right: HJD=2458166.352. Dashed curves: template  G2 V star,  $v\sin i = 66$ \kms. }
    \label{fig:SU_HA_NA}
\end{figure}

In SU Aur the H$\alpha$  profiles  show similar features of wind and accretion (Fig.~\ref{fig:SU_HA_NA}).  The central absorption at  about -40 \kms  is more narrow  and deep, and another broader absorption appear sometime at radial velocities of -100 to -250 \kms.   The same features are present in the \ion{Na}{I} D lines, which consist of broad photospheric absorption, narrow interstellar absorption, and the variable blue-shifted absorption indicating outflow.

\subsubsection{Spectral series of   RY Tau }
\label{sec:ry} 

\begin{figure}
\center\includegraphics[trim={-2cm 17.7cm 0cm 0cm},clip,width=1.16\columnwidth]{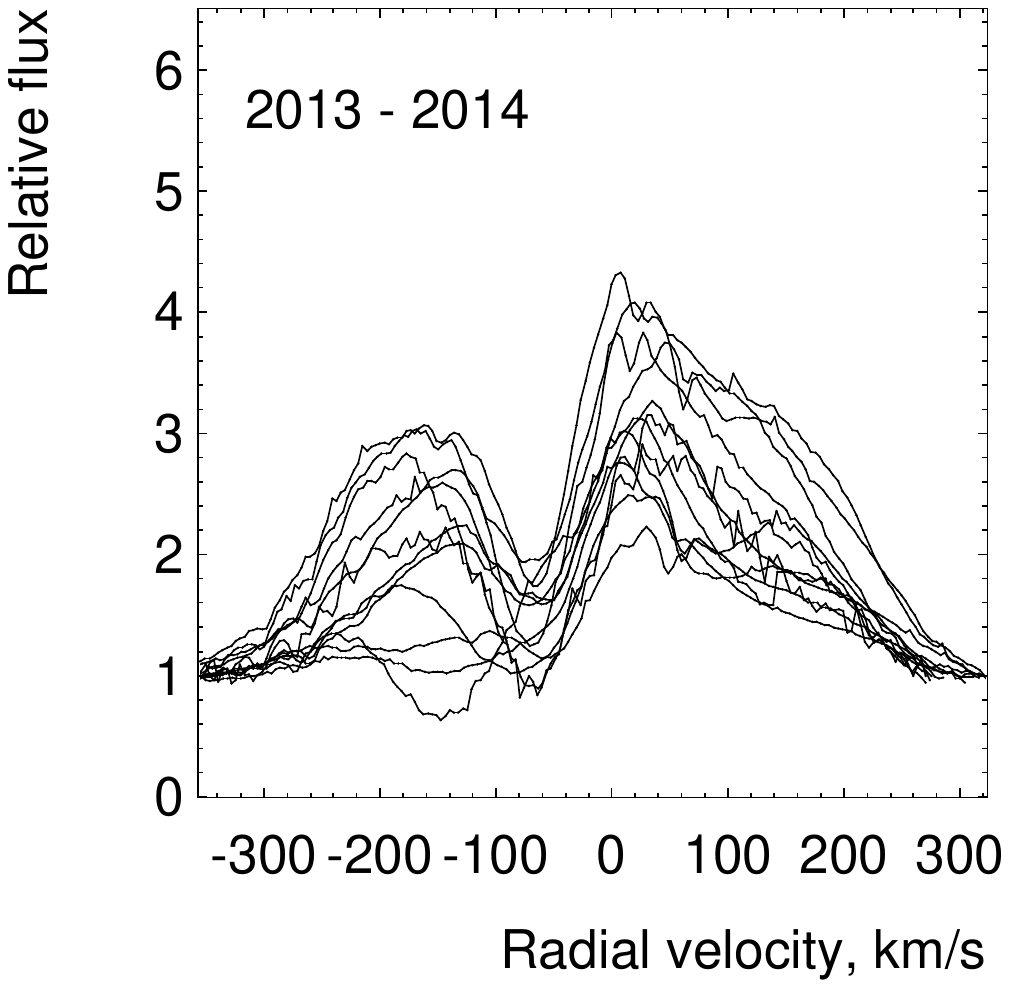}  
\vspace{-0.7cm}
\center\includegraphics[trim={-2cm 17.7cm 0cm 0cm},clip,width=1.16\columnwidth]{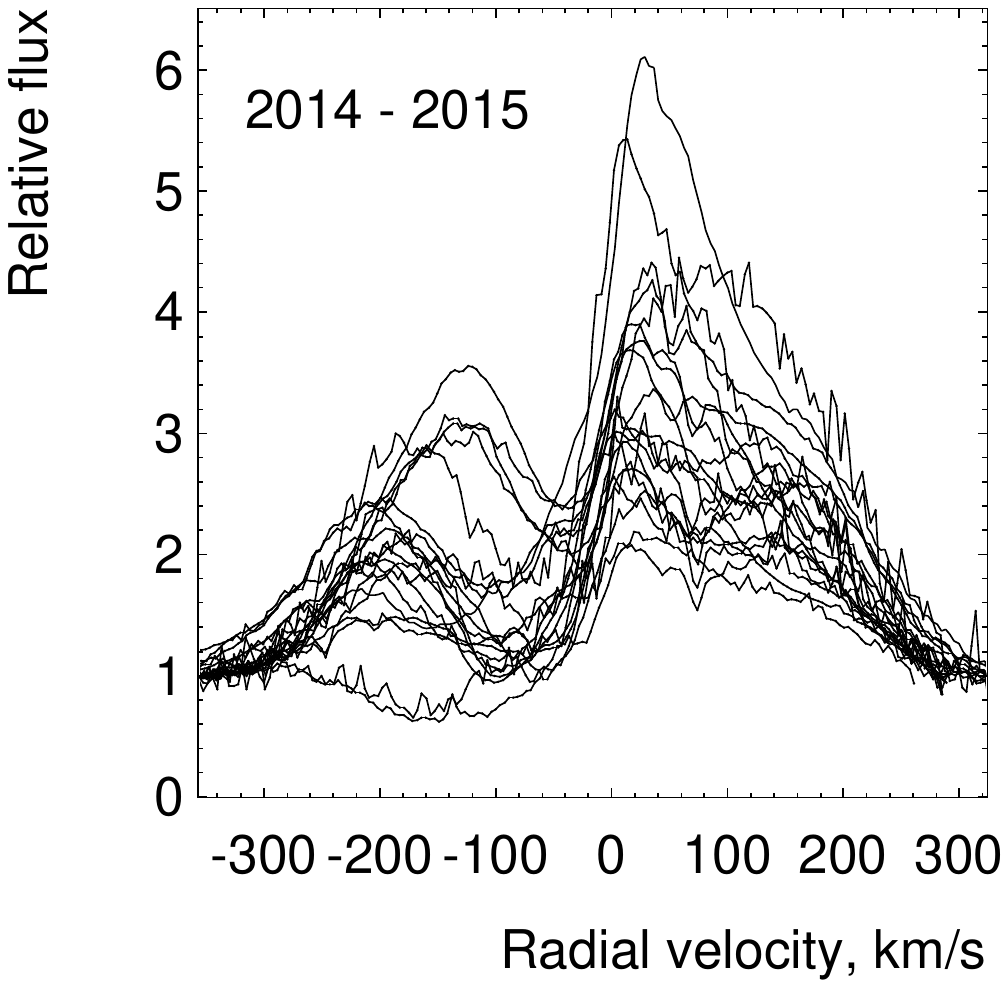}
\vspace{-0.7cm}
\center\includegraphics[trim={-2cm 17.7cm 0cm 0cm},clip,width=1.16\columnwidth]{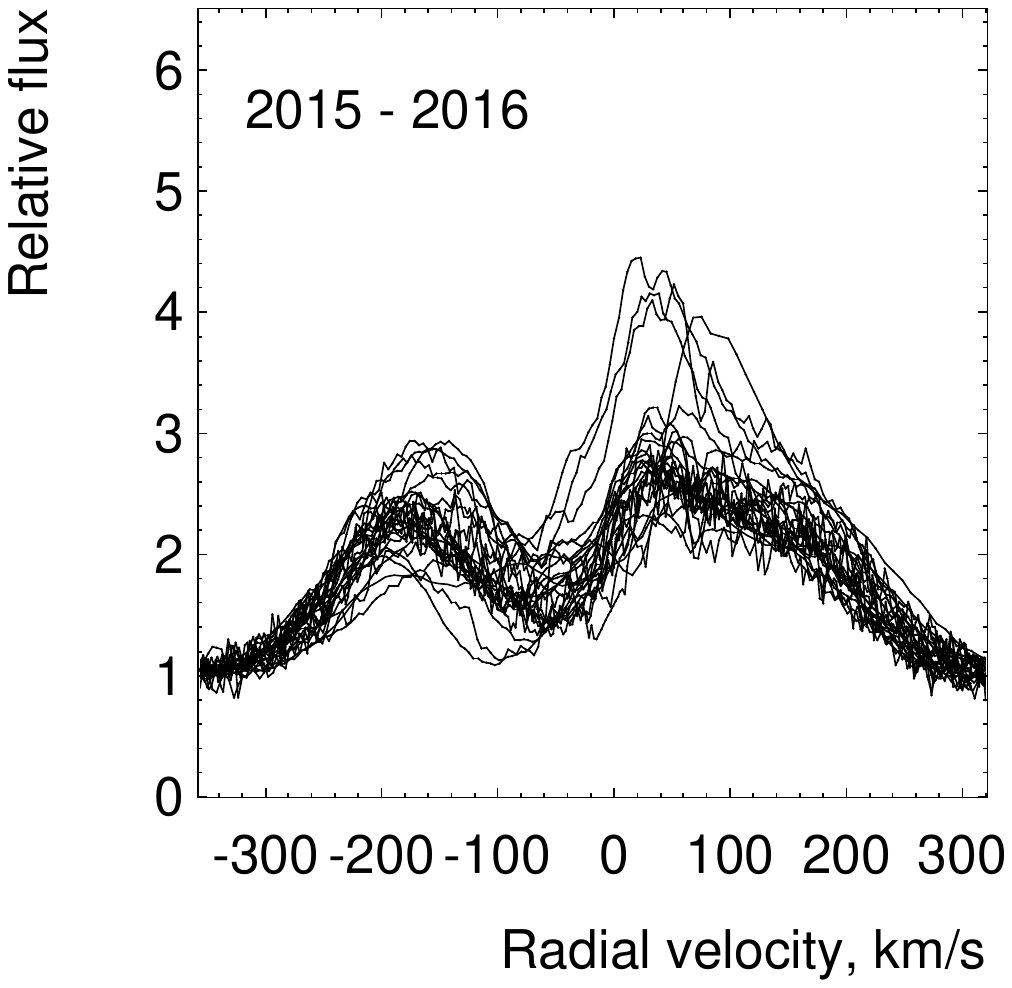}
\vspace{-0.7cm}
\center\includegraphics[trim={-2cm 17.7cm 0cm 0cm},clip,width=1.16\columnwidth]{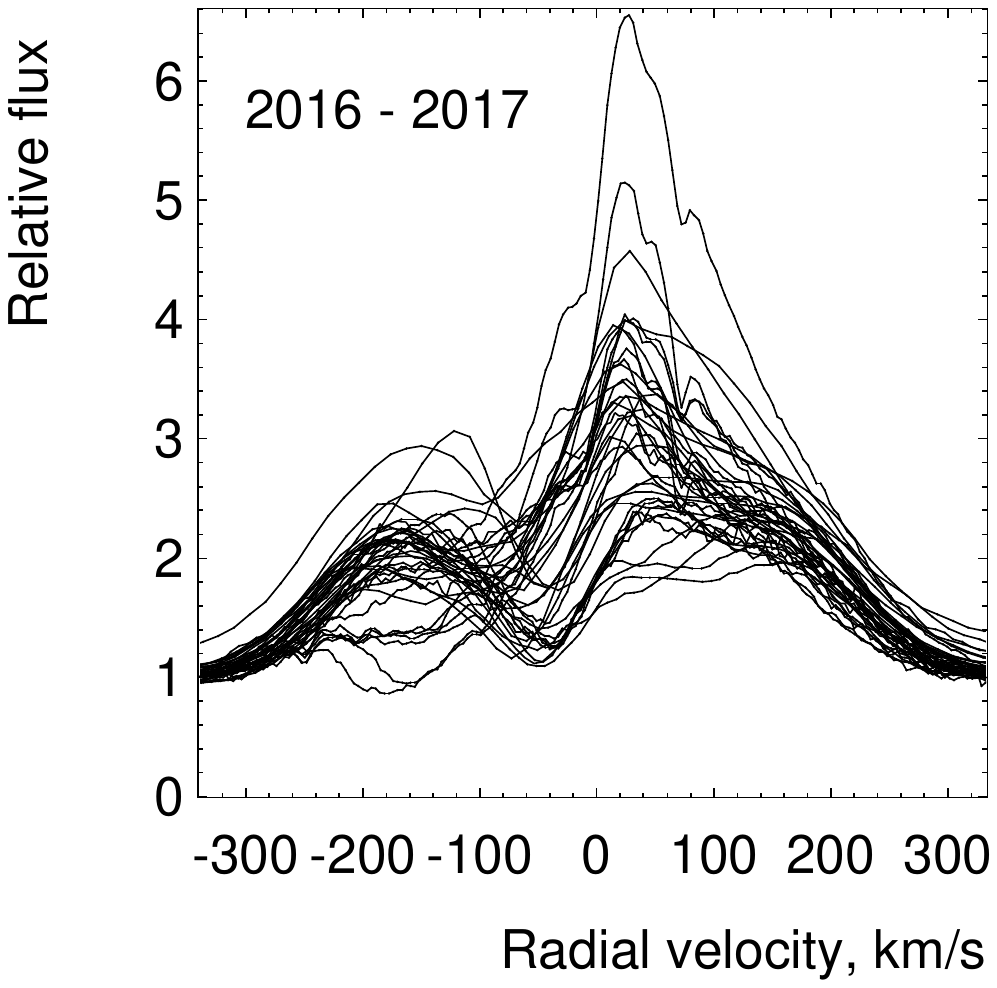}
\vspace{-0.7cm}
\center{\includegraphics[trim={-2cm 17.7cm 0cm 0cm},clip,width=1.16\columnwidth]{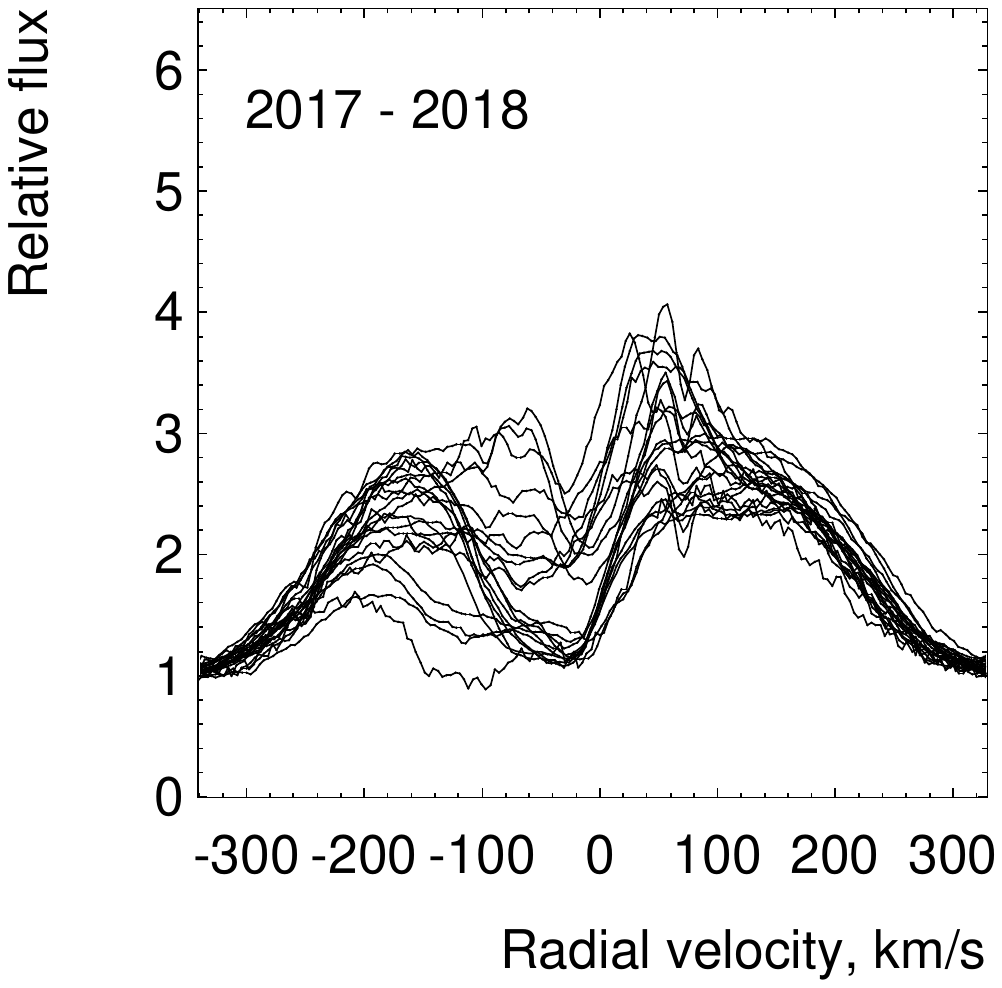}}
\caption{H$\alpha$  profile variation in RY Tau in five seasons  as indicated in the upper left corner of each panel.  The high activity in 2014-2015  was changed to a period of quiescence in 2015-2016.}
\label{fig:RY_5HAL}
\end{figure}

The H$\alpha$ emission is an indicator of MHD processes at the boundary between the stellar magnetosphere and the accretion disc.  The stellar brightness is related to the amount of dust around the star. In the following analysis we seek for a possible connection between variations in  H$\alpha$ emission and in the circumstellar extinction, i.e. the stellar brightness.

\begin{figure}
  \center{\includegraphics[trim={3.5cm 12cm 2.5cm 8cm},clip,width=1.1\columnwidth]{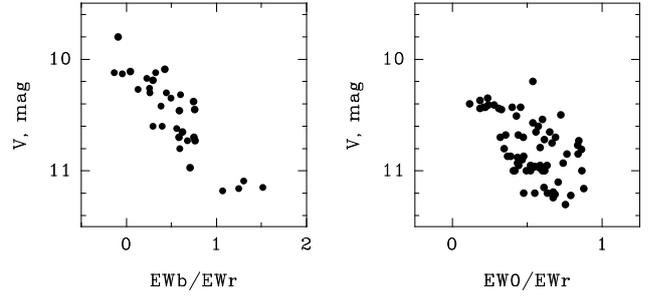}}
    \caption{Relation between the stellar brightness V and the H$\alpha$ line asymmetry. Left: data of 2013-2015,
right: data of 2016-2018. The star is brighter when the line asymmetry indicates stronger wind.}
    \label{fig:RY_V_EW}
\end{figure}

 The range of H$\alpha$ profile variability in RY Tau is shown in Fig.~\ref{fig:RY_5HAL}.
The most variable part of the H$\alpha$ profile is the central peak of emission at about +50\kms and the depression in the blue wing at about -100 to -200 \kms. In order to quantify the profile variability, we measured equivalent widths (EW) in three ranges of the velocity scale: EWb at -200 to -100 \kms, EW0 at -100 to 0  \kms, and  EWr at 0 to +100 \kms.   The ratio  EWb/EWr (or EW0/EWr) is a measure of the line asymmetry, caused mostly by the outflow.  Fig.~\ref{fig:RY_V_EW}  shows the relation between  the H$\alpha$ profile asymmetry and stellar brightness V. In the first  two seasons, when the line profile was most variable, there was a clear correlation: the blue wing of H$\alpha$ emission was depressed at the moments of high brightness of the star. Consequently, the circumstellar extinction was lower during times of enhanced outflow. In the last two seasons, when the activity resumed after the period of quiescence, the most variable part of the line profile shifted to lower velocities, and the ratio EW0/EWr showed a correlation with stellar brightness.

\subsubsection{Spectral series of SU Aur}
\label{sec:su} 

\begin{figure}
\center{\includegraphics[trim={-2cm 19cm 0cm 0cm},clip,width=1.5\columnwidth]{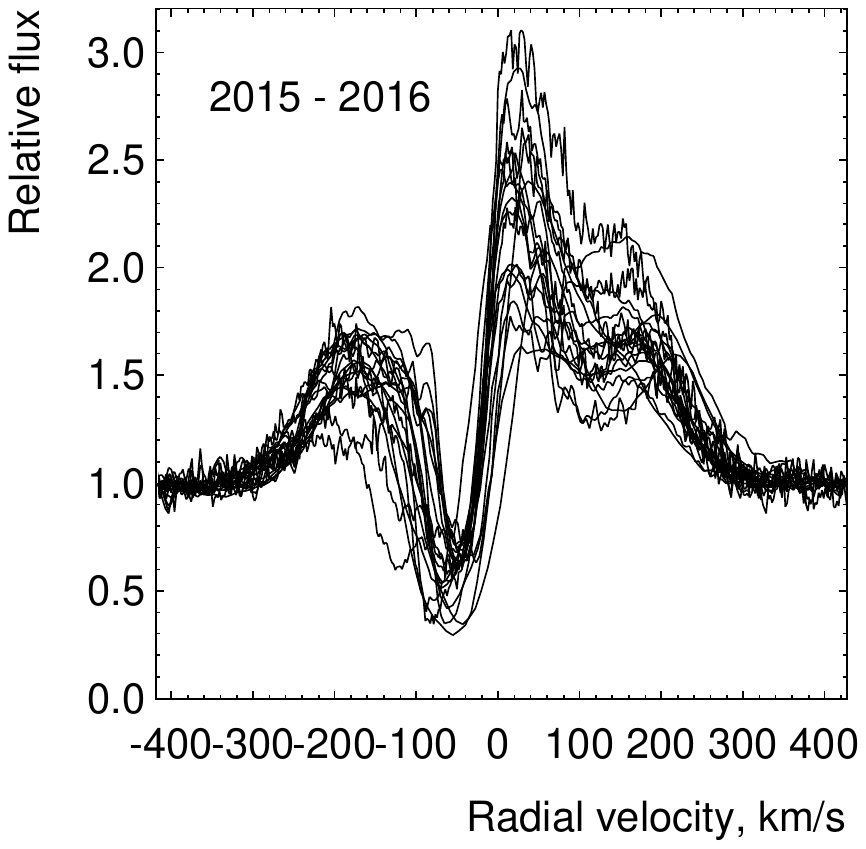}}
\vspace{-0.7cm}
\center{\includegraphics[trim={-2cm 19cm 0cm 0cm},clip,width=1.5\columnwidth]{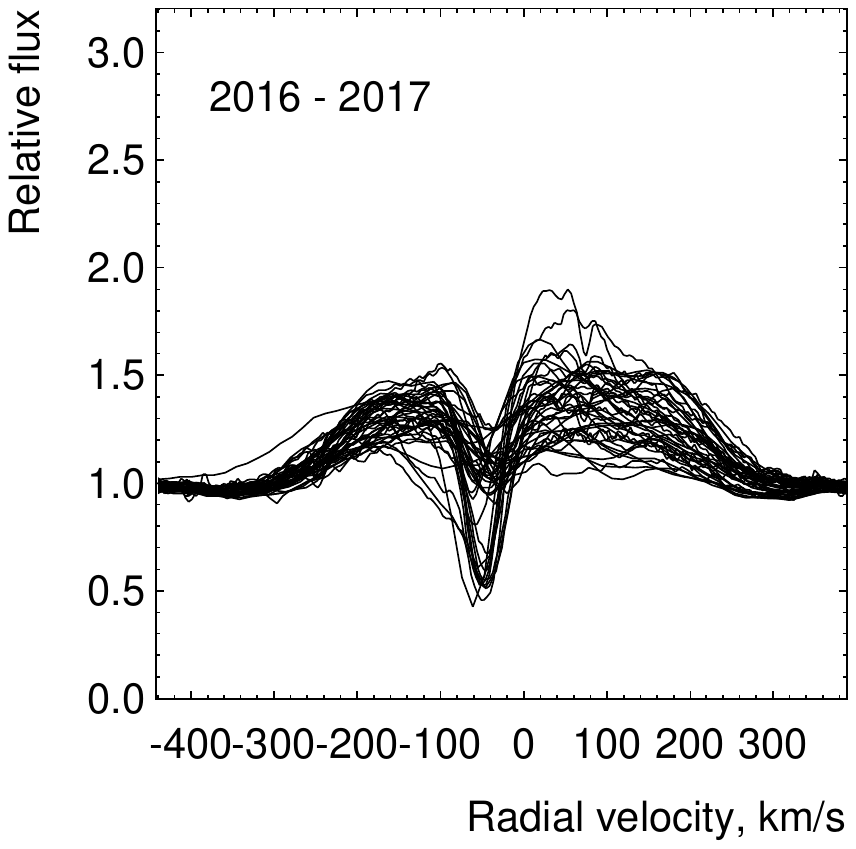}}
\vspace{-0.3cm}
\center{\includegraphics[trim={-3.1cm 19cm 0cm 0cm},clip,width=1.56\columnwidth]{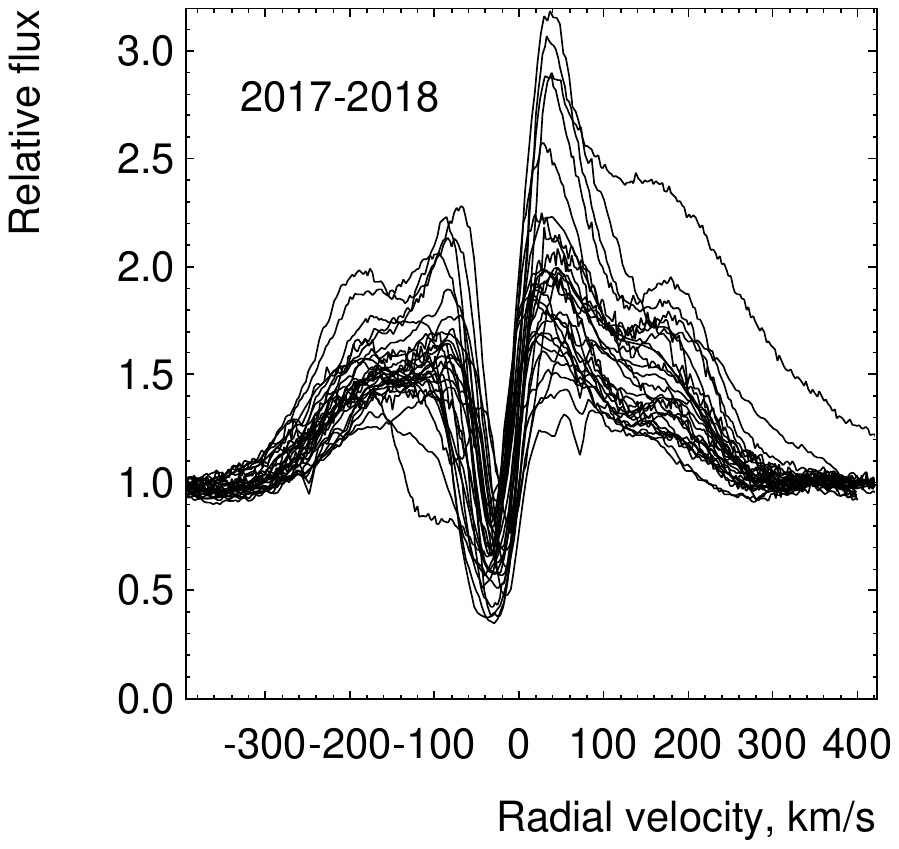}}
\caption{H$\alpha$  profile variation in SU Aur in three seasons. The season of observation is indicated in the upper left corner of each plot. }
    \label{fig:SU_3HAL}
\end{figure}

SU Aur was monitored only during three seasons. The range of H$\alpha$ profile variability is shown in Fig.~\ref{fig:SU_3HAL}. The line is broad with wings extending to radial velocities of about $\pm$ 400 \kms. On one occasion, the red emission wing extends to almost 600 \kms. Both wings vary in intensity, indicating irregular processes of accretion and outflows. There is a relatively stable narrow absorption at about -50 \kms. In terms of magnetospheric accretion and disc wind models, it may be identified with absorption in the disc wind or conical wind \citep{kurosawa2011,kurosawa2012}.

In the season of 2016-2017 the star showed an unusually low intensity in the H$\alpha$ emission line. The mean level of stellar brightness did not change considerably in that season, so it was a real decrease in the H$\alpha$ flux. In the \ion{Na}{I} D lines the absorption related to outflow was absent in some nights of that period. Interestingly, that in 2016-2017 the star also showed an unusual low amplitude in the brightness variations (Fig.~\ref{fig:SU_15_18}, middle panel). 
Unlike RY Tau,  there is  no correlation between the  H$\alpha$ line asymmetry and the stellar brightness in SU Aur. 


\subsubsection{H$\alpha$ flux variations}
\label{sec:Hflux} 

So far we analysed H$\alpha$ equivalent width and the line profile.
In the following discussion, an important parameter is the  flux radiated in H$\alpha$.
Our photometry enables  to transform the    equivalent
 width of H$\alpha$ into flux: $F = EW \times 10 ^{-0.4 \times(V-V_0)}$, where $V_0$  is a reference level  of stellar brightness, e.g. $V_0=10$ mag.  In this case the flux is 
expressed in units of the continuum flux density of a star with $V=10$ mag, which is $3.67 \times 10^{-13}$ erg cm$^{-2}$ s$^{-1}$ \AA$^{-1}$. The photometric R band is 
more appropriate for H$\alpha$ flux calibration,  but for some spectral observations only V is available from the AAVSO data. In RY Tau, the colour $(V-R)$ does not change 
considerably with brightness (see Fig.~\ref{fig:RY_color}). On the average, $(V-R) = 1.1 \pm 0.1$ mag. The use of  V magnitudes introduces a constant factor to the flux adding a 
relative error of about $10\%$. In our analysis, the absolute value of the flux is not critical.  

\begin{figure}
\center{    \includegraphics[trim={2cm 12cm 0cm 2cm},clip,width=1.2\columnwidth]{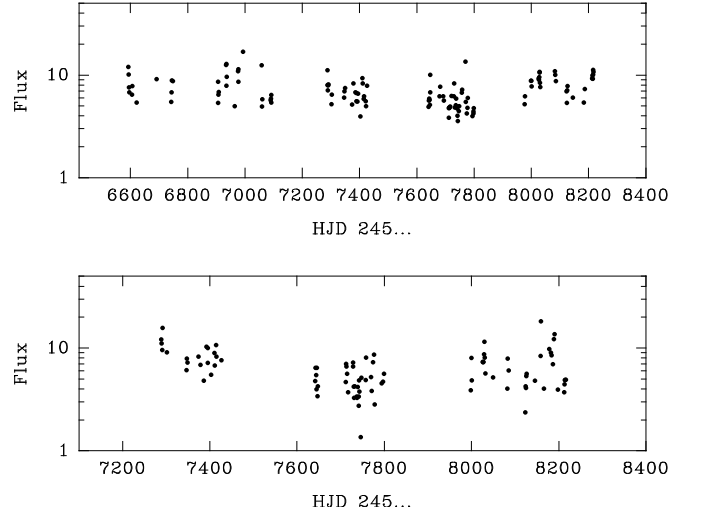}}
    \caption{Time variations in the absolute flux in H$\alpha$ emission line. Upper panel:  RY Tau, lower panel: SU Aur. Flux is in units of  $3.67 \times 10^{-13}$ erg cm$^{-2}$ s$^{-1}$ \AA$^{-1}$.  }
    \label{fig:Hal_flux}
\end{figure}

Time variations of the absolute flux in H$\alpha$ emission line in RY Tau and SU Aur are shown in  Fig.~\ref{fig:Hal_flux}.


\section{discussion}

The photometric and spectral properties of RY Tau and SU Aur, discussed in the previous sections, reveal similarities and differences between the two PMS stars. Stellar parameters are about the same, but RY Tau is younger and  more obscured by the circumstellar dust. 

The powerful outflows in cTTS are driven by accretion,  and the observed variations of the outflows are  related to the unstable MHD processes at the  boundary between the inner disc and stellar magnetosphere \citep{zanni2013}.  In the observed  H$\alpha$ profiles,  the most stable feature is the ''central'' absorption at -100 to -50 \kms. This absorption is related to an extended disc wind in the case of high  inclinations ($50^\circ-80^\circ$) of its rotational axis to the line of sight \citep{kurosawa2011}. The relative stability of this feature is due to a large area along the line of sight above the disc plane, where the absorption is formed. Alternatively, a similar type of profile can be formed in a conical wind starting near the interface of the magnetosphere and the accretion disc, when the stellar dipole magnetic field is compressed by the accretion disc into an X-wind like configuration \citep{kurosawa2012}.

More variable is the blue wing of H$\alpha$ emission at  -200 to -100 \kms. The amplitude of the flux variation in that region is large, including occasional appearance of the classical P Cyg type profile.  Such profiles indicate radially expanding outflows, which is drastically different from the disc wind. This fast expansion can be identified with a  Magnetospheric Ejection (ME).  The MEs appear when the inner disc comes closer to the star and the faster Keplerian rotation twists the stellar magnetosphere in azimuthal direction, which results in cyclically repeated openings and reconstructions of the magnetosphere \citep{goodson1997,zanni2013}.

The wind dynamics  and variations in the circumstellar  extinction are more evident in RY Tau. This may be a consequence of the higher accretion rate. The mass accretion rate can be estimated from the equivalent widths of the Hydrogen and Helium emission lines, using the empirical relations between line luminosity and mass accretion rate (e.g., \citealt{alcala2014}).
We derived the average accretion rates of our targets as: $\approx3.6\times10^{-8}$  \Msuny for RY Tau and $\approx4\times10^{-9}$ \Msuny for SU Aur. This is consistent with the previous determination of the mass accretion rates from optical-UV data on these stars \citep{calvet2004}.
 
In the following discussion we consider  two topics: 1) influence of the MEs on the dusty disc wind, and 2) the nature of  the gradual changes in the wind activity on a time scale of 1-2 years.

\subsection{ Interaction between the wind and the dusty environment}
 
The observed decrease of circumstellar extinction at the moments of the most intensive MEs ( Fig.~\ref{fig:RY_V_EW})  in RY Tau provides a possibility to localize the dust responsible for the extinction.
 Our observations showed that the characteristic radial velocity of the MEs is about 200 \kms, and the characteristic time scale of MEs is about two days 
\citep{babina2016} .  
Then, the typical distance from the expanding magnetosphere to the obscuring dust screen must be about 0.2 AU. 
{Assuming the magnetospheric radius of about 5 stellar radii (= 0.08 AU) we get the location of the dust screen at about 0.3 AU.  }
 This is near the inner edge of the dusty disc, where the dust temperature is high.  The SED in the NIR  (Fig.  \ref{fig:RY_SED}) shows radiation of dust at T=1500 K. Therefore, this hot dust can be identified as a cause of the variable circumstellar extinction. The star is seen through the dust screen, where the circumstellar extinction on a line of sight is changing, while the bulk radiation from the dust remains relatively constant. Similar effect was observed in another cTTS,  namely RW Aur. During a deep minimum of optical brightness of the star the NIR radiation increased, thus  indicating appearance of a hot dust  \citep{shenavrin2015}.

An accretion disc is a reservoir of dust. Coarse dust grains are concentrated at the disc plane, while small particles are present in the disc atmosphere and can be elevated from the disc plane by dynamical pressure of the disc wind \citep{safier1993}.   The disc wind is most intensive at the inner part of the disc, where the temperature is higher. Therefore, the most  dense dust screen is formed at the inner disc, near the dust sublimation distance and further out from the star.


The disc wind flows along the open magnetic field lines of the disc. 
 The observed correlation between the H$\alpha$ profile variations  and stellar brightness  (Fig.\ref{fig:RY_V_EW})  implies that
there  must be a physical mechanism of interaction between the MEs and the dusty disc wind close to the magnetosphere. As a tentative explanation
we suggest that a parcel of ionized gas, ejected from magnetosphere can affect the magnetic field of the  inner disc and thus temporarily change the disc wind flow on the line of sight.

 
In RY Tau this effect was more pronounced during the two seasons of maximal activity of magnetospheric ejections in 2013-2015.  Then the star entered  a  period of quiescence, when both the wind activity and the light variations became lower. In the last two seasons the effect appeared again. 
Since this effect has been observed repeatedly for several years, it can be considered as well established.
In SU Aur such a connection between the wind  and the circumstellar extinction  was not observed.

The colour-magnitude diagrams (Fig.\ref{fig:RY_color}) show that RY Tau is permanently hidden by the dust screen, so that the intensity of the light scattered on the dust particles is comparable to the intensity of the direct star light. 
Contrary to RY Tau,  SU Aur is at normal (high) brightness most of the time, but is occasionally obscured by circumstellar dust.  
This difference may be related to the age: RY Tau is younger and possess a more massive accretion disc \citep{akeson2005}. 

\subsection{Seasonal changes in outflow activity}

Our results show that there is a gradual change  of the outflow activity on time scales of a few years. In RY Tau we observed a period of activity  in the first two seasons, which was then replaced with a period  of quiescence between 2015 and 2016. Although the number of observations  in that season were large, the H$\alpha$ emission showed only a small amplitude of variability. A similar drop of activity was observed in SU Aur between 2016 and 2017.  Obviously, in the periods of quiescence the mechanism of ME did not work. 

\begin{figure}
\center{    \includegraphics[trim={2cm 16cm 0cm 3.5cm},clip,width=1.2\columnwidth]{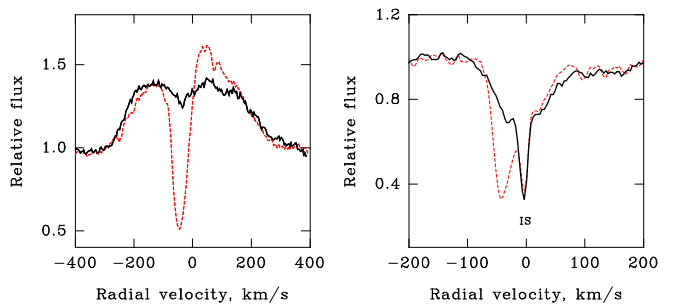}}
    \caption{Disappearance of the wind features in  SU Aur in December 2016. Left: H$\alpha$, right : \ion{Na}{I} D2. Dashed lines: typical profiles with blue-shifted wind absorptions. Solid lines: spectrum of December 05,  2016.  IS: interstellar absorption in the D2 line.}
    \label{fig:low_wind}
\end{figure}

Remarkable is the period of quiescence in SU Aur,  when the disc wind became very weak and almost absent. Fig.~\ref{fig:low_wind} shows that the usually strong ''central'' absorption in H$\alpha$ almost disappeared in December 2016. The corresponding absorption feature in the sodium doublet disappeared completely: the \ion{Na}{I} D2 line showed only the broad photospheric absorption and the narrow interstellar feature. The period  when the  wind was weak started in November 2016 and lasted to the end of the season. In the beginning of the next season (September 2017) the wind was active again. The minimal EW of H$\alpha$ reached 0.5 -- 1.0 \AA, and the corresponding  accretion rate  was less than 10$^{-9}$ \Msuny. Interestingly, that during the quiescence period SU Aur was also quiet photometrically, with a minimal circumstellar extinction and low variability within V= 9.2 -- 9.4 (See Fig.~\ref{fig:SU_15_18}, middle panel).

The  outflow is accretion driven, and the region of most unstable outflow is  the interface between the disc and stellar magnetosphere. 
Could the quiescence periods be a consequence of a  temporal lowering of accretion rate?
The absolute flux in H$\alpha$ line is often used as a measure of the mass accretion rate. 
Figure \ref{fig:Hal_flux} shows that the instant accretion rate is highly variable, while the average level does not change significantly from season to season. 
From the five seasons of our observations of RY Tau we do not see  a correlation between the wind activity and the average H$\alpha$ flux in a season. 
The quiescent period in 2015-2016 corresponds to an average level of the H$\alpha$ flux. 
In SU Aur we have only three seasons of observations, therefore it is difficult to make a conclusion.
The period of quiescence in 2016-2017 corresponds to a slightly lower H$\alpha$ flux level.

Another important parameter is the stellar magnetic field, which truncates the accretion disc. 
Zeeman Doppler Imaging of cTTS shows a multipolar structure of magnetic fields (\citealt{johnstone2014} and references therein).
Apart from the stellar mass and mass accretion rate, the truncation radius of accretion disc is primarily determined 
by the strength of the dipole component of the field.  If there is solar-like magnetic cycle in cTTS, one would expect cyclic variations 
of the disc truncation radius \citep{johnstone2014}. 

Typically, the truncation radius is within the radius of corotation.
In this case the inner disc rotates faster than the stellar magnetosphere, and the regime of MEs becames possible \citep{zanni2013}.
In case of a stronger dipole component of the stellar magnetic field, the disc is truncated outside the corotation radius and the propeller regime is activated \citep{romanova2009}.

The strength and topology of magnetic field in cTTS vary on a time scale of years  (e.g., \citealt{donati2011,donati2012}).                                                                                                                                 
We suggest that the observed changes of the outflow activity in our targets, including the periods of quiescence,  may be related to a slow variation of the 
global stellar magnetic fields, regardless  of the type of wind.   RY Tau and SU Aur are fast rotating stars with radiative cores and convective envelopes, 
so the generation of the stellar magnetic field may be similar to the solar one. 
Spectral monitoring of the H$\alpha$ profile variability in selected cTTS during several years could potentially reveal cycles of magnetic activity, if any.
\section{Conclusions}

1)  RY Tau and SU Aur have about the same stellar parameters, but different circumstellar environment. RY Tau is younger, with the higher accretion rate and a more massive accretion disc. In both stars, the light variability in the optical region is  mostly due to the circumstellar extinction.

2) The dusty screen responsible for the variable circumstellar extinction of RY Tau is located at the inner edge of the accretion disc: this is a dusty disc wind. The obscuring dust has a temperature of about 1500 K.  The circumstellar extinction in the line of sight is variable, while the NIR radiation from the bulk of the dust remains about the same. 

3) In RY Tau the dusty screen permanently obscures the star: the star is never seen free of foreground dust. Unlike RY Tau, SU Aur exhibits a moderate circumstellar extinction most of the time, with  rare cases of light drops due to more  extinction.

4) In RY Tau we discovered a new effect: during the events of enhanced outflow the cicumstellar extinction gets lower. 
This indicates that  outflows may affect the inner dusty wind.

5) In both RY Tau and SU Aur we detected periods of quiescence: lower amplitude variations in the H$\alpha$  line profiles. 
The quiescence periods lasted for about one year.

\section*{Acknowledgements}

This research was supported by the RFBR grant 16-02-00140.

Based partly on observations made with the Nordic Optical Telescope, operated by the Nordic Optical 
Telescope Scientific Association at the Observatorio del Roque de los Muchachos, La Palma, Spain, of 
the Instituto de Astrofisica de Canarias, using ALFOSC, which is provided by the Instituto de 
Astrofisica de Andalucia (IAA) under a joint agreement with the University of Copenhagen and NOTSA.

We  acknowledge support from a Visitors Program at the   
Department of Astronomy at Stockholm University.
JFG and RMGA were supported by Funda\c{c}\~ao para a Ci\^encia e a   
Tecnologia (FCT) through national funds (UID/FIS/04434/2013) and by   
FEDER through COMPETE2020 (POCI-01-0145-FEDER-007672). RMGA is   
supported by the fellowship PD/BD/113745/2015, under the FCT PD   
Program PhD::SPACE, funded by FCT (Portugal) and POPH/FSE (EC) and by   
CRUP through a cooperation program (PAULIF: TC-16/17).
DEM acknowledges his work as part of the research activity supported
by the National Astronomical Research Institute of Thailand
(NARIT), Ministry of Science and Technology of Thailand.
SYuG was supported in part by the Ministry of Education and Science of Russia
(the basic part of the Stateassignment, RK no. AAAA-A17-117030310283-7)
and by the Act no. 211 of the Government of the Russian Federation, 
agreement № 02.A03.21.0006.

We acknowledge with thanks the variable star observations from the   
AAVSO International Database contributed by observers worldwide and   
used in this research.

This work has made use of data from the European Space Agency (ESA)   
mission {\it Gaia} (\url{https://www.cosmos.esa.int/gaia}), processed   
by the {\it Gaia} Data Processing and Analysis Consortium   
(DPAC,\url{https://www.cosmos.esa.int/web/gaia/dpac/consortium}).   
Funding for the DPAC has been provided by national institutions, in   
particular the institutions participating in the {\it Gaia}   
Multilateral Agreement.

PPP thanks Marina Romanova for valuable comments.


\bibliographystyle{mnras}
\bibliography{my_bibliography} 


\newpage
\appendix
\section{Observation data}


\begin{table*}
\centering
\caption{RY Tau V-magnitudes for the dates of spectral observations. The Heliocentric Julian Date (HJD) in the first column is followed by the site of the observation (CAHA - Calar Alto Observatory, CrAO -  Crimean Astrophysical Observatory, NOT - Nordic Optical Telescope, TNO - Thai National Observatory), magnitude measured in the V-band, corresponding error and source (CAS - Crimean Astronomical Station, AAVSO - American Association of Variable Star Observers, Int - value interpolated from AAVSO and CrAO photometry).}
\label{tab:table02}
\begin{tabular}{lcccc} 
\hline	
\hline
HJD-2400000& Site & V      & V error & Source of V   \\
\hline
56592.444 & CrAO & 10.46 & 0.01 & CrAO \\
56593.438 & CrAO & 10.70 & 0.01 & CrAO \\
56594.340 & CrAO & 10.45 & 0.01 & CrAO \\
56595.306 & CrAO & 10.38 & 0.01 & CrAO \\
56605.436 & CrAO & 10.09 & 0.01 & CrAO \\
56606.442 & CrAO & 10.11 & 0.01 & CrAO \\
56621.318 & CrAO & 10.19 & 0.01 & CrAO \\
56691.389 & CrAO & 9.87 & 0.01 & AAVSO \\
56742.224 & CrAO & 10.68 & 0.10 & Int \\
56743.185 & CrAO & 10.79 & 0.10 & Int \\
56744.242 & CrAO & 10.91 & 0.10 & Int \\
56748.291 & CrAO & 10.97 & 0.01 & CrAO \\
56905.524 & CrAO & 10.29 & 0.01 & CrAO \\  
56906.520 & CrAO & 10.26 & 0.01 & CrAO \\
56907.504 & CrAO & 10.30 & 0.10 & Int \\
56908.509 & CrAO & 10.35 & 0.01 & CrAO \\
56933.513 & CrAO & 10.13 & 0.10 & Int \\
56934.475 & CrAO & 10.13 & 0.10 & Int \\
56935.416 & CrAO & 10.12 & 0.01 & CrAO \\
56936.415 & CrAO & 10.17 & 0.01 & CrAO \\
56964.464 & CrAO & 10.32 & 0.01 & CrAO \\
56975.442 & CrAO & 10.42 & 0.01 & CrAO \\
56976.453 & CrAO & 10.60 & 0.01 & CrAO \\
56977.468 & CrAO & 10.58 & 0.10 & Int \\
56993.262 & CrAO & 10.27 & 0.01 & CrAO \\
57058.188 & CrAO & 10.62 & 0.01 & CrAO \\
57059.231 & CrAO & 10.76 & 0.01 & AAVSO \\
57060.188 & CrAO & 10.78 & 0.10 & Int \\
57089.231 & CrAO & 11.09 & 0.01 & CrAO \\
57090.203 & CrAO & 11.18 & 0.01 & CrAO \\
57091.204 & CrAO & 11.16 & 0.02 & CrAO \\
57092.230 & CrAO & 11.15 & 0.08 & CrAO \\
57287.438 & CrAO & 10.61 & 0.01 & CrAO \\  
57288.532 & CrAO & 10.61 & 0.01 & CrAO \\
57289.438 & CrAO & 10.61 & 0.01 & CrAO \\
57290.511 & CrAO & 10.69 & 0.01 & CrAO \\
57291.444 & CrAO & 10.78 & 0.01 & CrAO \\
57301.430 & CrAO & 10.64 & 0.01 & CrAO \\
57302.482 & CrAO & 10.56 & 0.01 & AAVSO \\
57346.369 & CrAO & 10.75 & 0.01 & CrAO \\
57347.483 & CrAO & 10.68 & 0.01 & CrAO \\
57349.166 & TNO & 10.75 & 0.01 & AAVSO \\
57374.180 & CrAO & 10.74 & 0.01 & CrAO \\
57378.300 & CrAO & 10.93 & 0.01 & CrAO \\
57386.009 & TNO & 10.53 & \textless{}0.01 & AAVSO \\
57389.298 & CrAO & 10.80 & 0.01 & AAVSO \\
57389.413 & CAHA & 10.80 & 0.01 & AAVSO \\
57392.334 & CrAO & 10.78 & 0.01 & CrAO \\
57403.056 & TNO & 10.50 & 0.10 & Int \\
57410.370 & CAHA & 10.38 & \textless{}0.01 & AAVSO \\
57411.368 & CAHA & 10.40 & 0.10 & Int \\
57414.380 & CAHA & 10.52 & 0.01 & CrAO, CAS \\
57415.458 & CAHA & 10.59 & 0.10 & Int \\
57421.302 & CrAO & 11.04 & 0.01 & CrAO \\
57422.315 & CrAO & 11.04 & 0.10 & Int \\
57426.327 & CrAO & 10.98 & 0.01 & CrAO \\

57641.426 & CrAO & 10.96 & 0.01 & CrAO, CAS \\   
57642.442 & CrAO & 10.95 & 0.01 & CrAO \\
57643.418 & CrAO & 10.91 & 0.01 & CrAO, CAS \\
57644.402 & CrAO & 10.88 & 0.01 & CrAO \\
57645.393 & CrAO & 11.01 & 0.01 & CrAO \\
57646.416 & CrAO & 10.96 & 0.01 & CrAO \\
&&&&\\
&&&&\\
&&&&\\

\end{tabular}
\quad
\begin{tabular}{ccccc}
\hline
\hline
HJD-2400000 & Site & V      & V error & Source of V   \\
\hline
57647.424 & CrAO & 10.97 & 0.01 & CrAO \\
57679.618 & NOT & 10.93 & \textless{}0.01 & AAVSO \\
57681.607 & NOT & 11.07 & 0.10 & Int \\
57691.681 & NOT & 11.20 & 0.10 & Int \\
57694.607 & NOT & 11.06 & 0.02 & AAVSO \\
57711.424 & CrAO & 11.31 & 0.01 & CrAO \\
57712.387 & CrAO & 11.22 & 0.01 & CrAO \\
57713.442 & CrAO & 11.16 & 0.01 & CrAO \\
57714.418 & CrAO & 11.27 & 0.01 & CrAO, CAS \\
57716.650 & NOT & 11.04 & 0.02 & AAVSO \\
57721.578 & NOT & 11.02 & 0.10 & Int \\
57728.594 & NOT & 11.16 & 0.01 & CrAO, CAS \\
57730.459 & NOT & 11.15 & 0.01 & CrAO \\
57732.546 & NOT & 11.15 & 0.10 & Int \\
57736.650 & NOT & 10.950 & 0.01 & CAS \\
57737.553 & NOT & 10.96 & 0.01 & CAS \\
57741.072 & TNO & 10.95 & 0.01 & CrAO, CAS \\
57742.091 & TNO & 10.90 & 0.10 & Int \\
57743.237 & TNO & 10.83 & 0.10 & Int \\
57746.144 & TNO & 10.74 & 0.01 & CrAO, CAS \\
57747.137 & TNO & 10.71 & 0.01 & CrAO, CAS \\
57752.570 & CAHA & 10.70 & 0.10 & Int \\
57753.464 & CAHA & 10.69 & 0.10 & Int \\
57754.497 & CAHA & 10.68 & 0.04 & AAVSO \\
57757.407 & NOT & 10.63 & 0.01 &  CAS \\
57758.545 & NOT & 10.64 & \textless{}0.01 & AAVSO \\
57769.523 & NOT & 11.11 & 0.10 & Int \\
57771.485 & NOT & 10.85 & 0.10 & Int \\
57774.159 & CrAO & 10.99 & 0.01 & CrAO \\
57776.205 & CrAO & 11.00 & 0.10 & Int \\
57778.447 & NOT & 10.95 & 0.10 & Int \\
57794.194 & CrAO & 10.93 & 0.01 & CrAO \\
57797.186 & CrAO & 10.85 & 0.01 & CrAO, CAS \\
57798.200 & CrAO & 10.87 & 0.10 & Int \\
57799.180 & CrAO & 10.87 & 0.10 & Int \\
57998.442 & CrAO & 10.82 & 0.01 & CrAO \\
57999.442 & CrAO & 10.85 & 0.01 & CrAO \\
58000.445 & CrAO & 10.72 & 0.01 & CrAO \\
58025.441 & CrAO & 10.45 & 0.01 & CrAO \\
58026.483 & CrAO & 10.42 & 0.10 & Int \\
58027.476 & CrAO & 10.40 & 0.10 & Int \\
58028.475 & CrAO & 10.38 & 0.10 & Int \\
58029.435 & CrAO & 10.37 & 0.01 & CAS \\
58030.438 & CrAO & 10.43 & 0.01 & CAS \\
58031.438 & CrAO & 10.43 & 0.01 & CAS \\
58082.248 & CrAO & 10.45 & 0.01 & CrAO \\
58083.247 & CrAO & 10.51 & 0.01 & CrAO \\
58085.234 & CrAO & 10.83 & 0.10 & Int \\
58123.148 & CrAO & 10.46 & 0.01 & CAS \\
58124.275 & CrAO & 10.37 & 0.10 & Int \\
58125.149 & CrAO & 10.34 & 0.01 & CrAO \\
58126.144 & CrAO & 10.41 & 0.01 & CrAO, CAS \\
58145.229 & CrAO & 11.18 & 0.01 & CrAO, CAS \\
58183.200 & CrAO & 10.55 & 0.01 & CrAO \\
58186.251 & CrAO & - & - & - \\
58212.224 & CrAO & 10.54 & 0.01 & CrAO \\
58213.225 & CrAO & 10.57 & 0.01 & CrAO \\
58214.217 & CrAO & 10.65 & 0.10 & Int \\
58215.220 & CrAO & 10.73 & 0.01 & CrAO \\
58216.215 & CrAO & 10.77 & 0.01 & CrAO \\
58217.219 & CrAO & 10.79 & 0.01 & CrAO\\  
&&&&\\
&&&&\\
&&&&\\
&&&&\\
\end{tabular}
\end{table*}

\begin{table*}
\centering
\caption{SU Aur V-magnitudes for the dates of spectral observations. The Heliocentric Julian Date (HJD) in the first column is followed by the site of the observation (CAHA - Calar Alto Observatory, CrAO -  Crimean Astrophysical Observatory, NOT - Nordic Optical Telescope, TNO - Thai National Observatory, UrFU - Kourovka Astronomical Observatory), magnitude measured in the V-band, corresponding error and source (CAS - Crimean Astronomical Station, AAVSO - American Association of Variable Star Observers, Int - value interpolated from AAVSO and CrAO photometry).}
\label{tab:table03}
\begin{tabular}{lcccc} 
\hline
\hline
HJD-2400000 & Site & V      & V error & Source of V   \\
\hline
57288.462 & CrAO & 9.34 & 0.01 & CrAO \\   
57289.503 & CrAO & 9.39 & 0.01 & CrAO \\
57290.490 & CrAO & 9.39 & 0.01 & CrAO \\
57291.502 & CrAO & 9.33 & 0.01 & CrAO \\
57301.500 & CrAO & 9.83 & 0.01 & CrAO \\
57346.440 & CrAO & 9.50 & 0.01 & CrAO \\
57347.510 & CrAO & 9.45 & 0.01 & CrAO \\
57349.190 & TNO & 9.46 & 0.10 & Int \\
57374.225 & CrAO & 9.43 & 0.01 & CrAO \\
57378.333 & CrAO & 9.44 & 0.01 & CrAO \\
57386.090 & TNO & 9.40 & 0.10 & Int \\
57392.351 & CrAO & 9.41 & 0.10 & Int \\
57395.086 & TNO & 9.50 & 0.01 & AAVSO \\
57395.406 & CAHA & 9.50 & 0.01 & AAVSO \\
57403.076 & TNO & 9.69 & \textless{}0.01 & AAVSO \\
57410.402 & CAHA & 9.72 & 0.01 & AAVSO \\
57411.401 & CAHA & 9.61 & 0.10 & Int \\
57414.412 & CAHA & 9.80 & 0.01 & CrAO \\
57415.489 & CAHA & 9.73 & 0.10 & Int \\
57426.373 & CrAO & 9.53 & 0.01 & CrAO \\
57641.536 & CrAO & 9.42 & 0.01 & CrAO \\  
57642.517 & CrAO & 9.52 & 0.01 & CrAO \\
57643.524 & CrAO & 9.43 & 0.01 & CrAO \\
57644.467 & CrAO & 9.41 & 0.01 & CrAO \\
57645.460 & CrAO & 9.39 & 0.01 & CrAO \\
57646.481 & CrAO & 9.38 & 0.01 & CrAO \\
57646.689 & NOT & 9.38 & 0.01 & CrAO \\
57647.490 & CrAO & 9.52 & 0.01 & CrAO \\
57679.623 & NOT & 9.42 & \textless{}0.01 & AAVSO \\
57681.613 & NOT & 9.51 & 0.01 & AAVSO \\
57691.688 & NOT & 9.35 & 0.10 & Int \\
57694.614 & NOT & 9.33 & 0.10 & Int \\
57711.499 & CrAO & 9.30 & 0.01 & CrAO \\
57712.584 & CrAO & 9.24 & 0.01 & CrAO \\
57713.599 & CrAO & 9.28 & 0.01 & CrAO \\
57714.486 & CrAO & 9.26 & 0.01 & CrAO \\
57716.662 & NOT & 9.22 & 0.10 & Int \\
57721.584 & NOT & 9.20 & 0.10 & Int \\
57728.213 & CrAO & 9.18 & 0.01 & CrAO \\
57728.600 & NOT & 9.18 & 0.01 & CrAO \\
57730.237 & CrAO & 9.20 & 0.01 & CrAO \\
57730.465 & NOT & 9.20 & 0.01 & CrAO \\
57732.643 & NOT & 9.19 & 0.10 & Int \\
57735.584 & CAHA & 9.17 & 0.10 & Int \\
57736.656 & NOT & 9.16 & 0.10 & Int \\
57737.560 & NOT & 9.16 & 0.10 & Int \\
57739.388 & CrAO & 9.15 & 0.01 & CrAO \\
57741.072 & TNO & 9.19 & 0.01 & CrAO \\
57742.091 & TNO & 9.20 & 0.10 & Int \\
57743.237 & TNO & 9.21 & 0.10 & Int \\
57746.144 & TNO & 9.21 & 0.01 & CrAO \\
57747.135 & TNO & 9.18 & 0.01 & CrAO \\
57752.617 & CAHA & 9.20 & 0.10 & Int \\
57753.509 & CAHA & 9.21 & 0.10 & Int \\
57754.547 & CAHA & 9.21 & 0.10 & Int \\
57757.413 & NOT & 9.21 & 0.01 & CrAO \\
57758.551 & NOT & 9.22 & 0.10 & Int \\
57769.537 & NOT & 9.22 & 0.10 & Int \\
57771.491 & NOT & 9.21 & 0.10 & Int \\
57774.232 & CrAO & 9.21 & 0.01 & CrAO \\  
57776.273 & CrAO & 9.22 & 0.01 & CrAO \\
\end{tabular}
\quad
\begin{tabular}{ccccc}
\hline
\hline
HJD-2400000 & Site & V      & V error & Source of V   \\
\hline
57778.455 & NOT & 9.22 & 0.10 & Int \\
57794.263 & CrAO & 9.32 & 0.01 & CrAO \\
57797.259 & CrAO & 9.40 & 0.01 & CrAO \\
57799.248 & CrAO & 9.44 & 0.10 & Int \\
57998.513 & CrAO & 9.40 & 0.01 & CrAO \\
57999.509 & CrAO & 9.30 & 0.01 & CrAO \\
58000.510 & CrAO & 9.39 & 0.01 & CrAO \\
58025.536 & CrAO & 9.41 & 0.01 & CrAO \\
58026.549 & CrAO & 9.40 & 0.10 & Int \\
58027.542 & CrAO & 9.40 & 0.10 & Int \\
58028.540 & CrAO & 9.40 & 0.10 & Int \\
58029.500 & CrAO & 9.40 & 0.10 & Int \\
58030.503 & CrAO & 9.43 & 0.01 & CAS \\
58031.502 & CrAO & 9.39 & 0.01 & CAS \\
58049.473 & UrFU & 9.60 & 0.01 & AAVSO \\
58082.322 & CrAO & 9.50 & 0.01 & CrAO \\
58083.312 & CrAO & 9.46 & 0.01 & CrAO \\
58085.302 & CrAO & 9.46 & 0.10 & Int \\
58123.193 & CrAO & 9.74 & 0.01 & CAS \\
58124.482 & UrFU & 9.98 & 0.01 & AAVSO \\
58125.216 & CrAO & 10.01 & 0.01 & CrAO \\
58125.404 & UrFU & 10.01 & 0.01 & CAS\\
58126.210 & CrAO & 9.88 & 0.01 & CrAO \\
58145.165 & CrAO & 10.76 & 0.01 & CAS \\
58158.293 & UrFU & 9.60 & 0.01 & AAVSO \\
58159.324 & UrFU & 9.66 & 0.01 & AAVSO \\
58162.338 & NOT & 9.62 & 0.10 & Int \\
58182.355 & CrAO & 9.45 & 0.01 & CrAO \\
58183.266 & CrAO & 9.47 & 0.01 & CrAO \\
58186.225 & CrAO & 9.62 & 0.10 & Int \\           
58212.262 & CrAO & 10.43 & 0.01 & CrAO \\
58213.270 & CrAO & 10.74 & 0.01 & CrAO \\
58214.242 & CrAO & 10.27 & 0.07 & AAVSO \\
58215.265 & CrAO & 10.23 & 0.01 & CrAO \\
58216.260 & CrAO & 10.01 & 0.01 & CrAO\\  
&&&&\\
&&&&\\
&&&&\\
&&&&\\
&&&&\\
&&&&\\
&&&&\\
&&&&\\
&&&&\\
&&&&\\
&&&&\\
&&&&\\
&&&&\\
&&&&\\
&&&&\\
&&&&\\
&&&&\\
&&&&\\
&&&&\\
&&&&\\
&&&&\\
&&&&\\
&&&&\\
&&&&\\
&&&&\\
&&&&\\
\end{tabular}
\end{table*}

\begin{table*}
\centering
\caption{RY Tau photometry. The values corresponding to the UBVRI photometry, taken between 1981 and 1997, are from the Majdanak archive \citep{grankin2007}. All the JHKLM magnitudes were obtained at the Crimean Astronomical Station (CAS). The BVRI data correspondent to the period 2013-2018 were obtained at  CrAO.}
\label{tab:table04}
\begin{tabular}{ccccccccccc} 
\hline
\hline
HJD-2400000& U      & B      & V      & R      & I     & J     & H     & K     & L     & M     \\
\hline
44888.400 & ...   & ...   & ...   & ...   & ...  & 7.80 & 6.65 & 5.55 & ...  & ...    \\
44892.400 & ...   & ...   & ...   & ...   & ...  & 7.73 & 6.60 & 5.51 & ...  & ...    \\
44902.600 & ...   & ...   & ...   & ...   & ...  & 7.78 & ...    & 5.61 & ...  & ...  \\
46692.400 & 11.97 & 11.40 & 10.24 & 9.02  & ...  & 7.26 & 6.17 & 5.26 & 4.16 & ...  \\
46694.386 & 11.90 & 11.29 & 10.15 & 8.95  & ...  & 7.34 & 6.19 & 5.25 & 4.08 & 3.74 \\
47053.500 & ...   & ...   & ...   & ...   & ...  & 7.02 & 6.05 & 5.24 & 4.07 & 3.68 \\
47126.500 & ...   & ...   & ...   & ...   & ...  & 7.26 & 6.24 & 5.35 & ...  & 3.65 \\
47569.300 & ...   & ...   & ...   & ...   & ...  & 6.99 & 5.99 & 5.32 & 4.17 & 3.80 \\
47825.369 & 11.81 & 11.20 & 10.09 & 8.96  & ...  & 7.15 & 6.12 & 5.26 & 4.09 & 3.70 \\
50430.400 & ...   & ...   & ...   & ...   & ...  & 6.95 & 6.03 & 5.28 & 4.25 & ...  \\
50484.300 & ...   & ...   & ...   & ...   & ...  & 6.88 & 5.92 & 5.39 & 4.21 & 3.82 \\
57289.431 & ...   & 11.66 & 10.61 & 9.58  & 8.70 & 7.37 & 6.37 & 5.41 & ...  & ...  \\
57291.500 & ...   & 11.84 & 10.78 & 9.72  & 8.80 & 7.47 & 6.40 & 5.44 & 4.23 & 3.88 \\
57385.300 & ...   & ...   & ...   & ...   & ...  & 7.34 & 6.37 & 5.48 & 4.30 & 4.03 \\
57391.291 & ...   & 11.77 & 10.73 & 9.66  & 8.76 & 7.36 & 6.39 & 5.47 & 4.31 & 4.09 \\
57408.263 & ...   & 11.44 & 10.42 & 9.38  & 8.51 & 7.22 & 6.26 & 5.38 & 4.16 & 3.84 \\
57427.300 & ...   & 11.93 & 11.01 & 10.06 & 9.16 & 7.52 & 6.39 & 5.41 & 4.14 & 3.89 \\
57444.193 & ...   & 11.99 & 11.04 & 10.06 & 9.17 & 7.61 & 6.42 & 5.55 & 4.23 & 3.98 \\
57453.200 & ...   & 11.89 & 11.00 & 10.04 & 9.20 & 7.60 & 6.45 & 5.41 & 4.16 & 3.91 \\
57620.486 & ...   & 11.82 & 10.86 & 9.85  & 8.97 & 7.51 & 6.46 & 5.50 & 4.26 & 4.07 \\
57623.500 & ...   & 11.84 & 10.87 & 9.85  & 8.95 & 7.51 & 6.45 & 5.46 & 4.20 & 4.11 \\
57635.464 & ...   & 11.96 & 10.91 & 9.83  & 8.94 & 7.58 & 6.54 & 5.56 & 4.30 & 3.96 \\
57638.500 & ...   & ...   & ...   & ...   & ...  & 7.61 & 6.52 & 5.53 & 4.25 & 3.92 \\
57641.488 & ...   & 11.99 & 10.96 & 9.90  & 8.99 & 7.60 & 6.52 & 5.52 & 4.28 & 3.94 \\
57648.492 & ...   & 11.91 & 10.95 & 9.93  & 9.09 & 7.68 & 6.59 & 5.57 & 4.27 & 3.92 \\
57680.478 & ...   & 11.93 & 10.94 & 9.86  & 8.92 & 7.43 & 6.36 & 5.38 & 4.19 & 3.98 \\
57704.400 & ...   & 12.08 & 11.21 & 10.15 & 9.27 & 7.68 & 6.55 & 5.50 & 4.19 & 4.03 \\
57730.332 & ...   & 12.21 & 11.15 & 10.06 & 9.13 & 7.61 & 6.51 & 5.44 & 4.17 & 3.68 \\
57745.300 & ...   & 11.80 & 10.68 & 9.50  & 8.57 & 7.30 & 6.28 & 5.39 & 4.11 & 3.80 \\
57746.216 & ...   & 11.86 & 10.72 & 9.56  & 8.63 & 7.28 & 6.30 & 5.35 & 4.06 & 3.80 \\
57767.208 & ...   & 12.21 & 11.12 & 9.97  & 9.02 & 7.54 & 6.43 & 5.44 & 4.15 & 3.98 \\
57774.297 & ...   & 12.12 & 10.99 & ...   & 8.93 & 7.45 & 6.37 & 5.39 & 4.17 & 3.88 \\
57777.200 & ...   & ...   & ...   & ...   & ...  & 7.53 & 6.42 & 5.41 & 4.15 & 4.01 \\
57784.185 & ...   & 11.90 & 10.87 & 9.80  & 8.90 & 7.45 & 6.42 & 5.44 & 4.24 & 3.97 \\
57785.272 & ...   & 11.95 & 10.85 & ...   & ...  & 7.41 & 6.38 & 5.44 & 4.20 & 4.02 \\
57786.162 & ...   & 11.92 & 10.86 & 9.75  & 8.83 & 7.39 & 6.38 & 5.45 & 4.20 & 3.99 \\
57795.200 & ...   & 11.93 & 10.90 & 9.80  & 8.85 & 7.33 & 6.31 & 5.37 & 4.13 & 3.88 \\
57801.278 & ...   & 11.88 & 10.89 & 9.83  & 8.93 & 7.47 & 6.44 & 5.45 & 4.20 & 3.78 \\
57803.189 & ...   & 11.86 & 10.88 & 9.88  & 9.05 & 7.56 & 6.48 & 5.51 & 4.20 & 3.88 \\
57812.200 & ...   & 11.84 & 10.79 & 9.69  & 8.76 & 7.45 & 6.47 & 5.54 & 4.28 & 3.97 \\
57813.226 & ...   & 11.84 & 10.78 & 9.68  & 8.76 & 7.39 & 6.38 & 5.52 & 4.21 & 3.94 \\
57817.202 & ...   & ...   & ...   & ...   & ...  & 7.34 & 6.34 & 5.46 & 4.23 & 3.94 \\
57998.512 & ...   & 11.85 & 10.81 & 9.67  & 8.76 & 7.42 & 6.42 & 5.49 & 4.31 & 3.95 \\
58006.541 & ...   & ...   & ...   & ...   & ...  & 7.36 & 6.37 & 5.43 & 4.26 & 4.06 \\
58026.513 & ...   & ...   & ...   & ...   & ...  & 7.20 & 6.28 & 5.42 & 4.20 & 3.88 \\
58030.448 & ...   & ...   & ...   & ...   & ...  & 7.17 & 6.25 & 5.39 & 4.23 & 3.89 \\
58038.489 & ...   & ...   & ...   & ...   & ...  & 7.24 & 6.32 & 5.43 & 4.27 & 3.97 \\
58066.397 & ...   & ...   & ...   & ...   & ...  & 7.03 & 6.17 & 5.36 & 4.18 & 3.96 \\
58096.465 & ...   & 11.43 & 10.40 & 9.29  & 8.44 & 7.22 & 6.32 & 5.44 & 4.20 & 3.95 \\
58100.320 & ...   & ...   & ...   & ...   & ...  & 7.30 & 6.36 & 5.46 & 4.23 & 3.96 \\
58114.436 & ...   & 11.78 & 10.73 & 9.65  & 8.75 & 7.29 & 6.34 & 5.38 & 4.10 & 3.62 \\
58116.293 & ...   & ...   & ...   & ...   & ...  & 7.43 & 6.41 & 5.44 & 4.20 & 3.93 \\
58120.251 & ...   & ...   & ...   & ...   & ...  & 7.23 & 6.25 & 5.36 & 4.19 & 3.90 \\
58125.437 & ...   & 11.44 & 10.34 & 9.23  & 8.34 & 7.10 & 6.19 & 5.34 & 4.21 & 3.95 \\
58126.210 & ...   & 11.54 & 10.41 & 9.29  & 8.39 & 7.14 & 6.23 & 5.36 & 4.19 & 3.91 \\
58143.219 & ...   & ...   & ...   & ...   & ...  & 7.46 & 6.47 & 5.51 & 4.22 & 3.84 \\
58145.209 & ...   & ...   & ...   & ...   & ...  & 7.46 & 6.45 & 5.48 & 4.28 & 4.03 \\
58151.178 & ...   & ...   & ...   & ...   & ...  & 7.31 & 6.34 & 5.44 & 4.24 & 4.00 \\
\hline
\end{tabular}
\end{table*}

\begin{table*}
\centering
\caption{SU Aur photometry. All the JHKLM magnitudes were obtained at the Crimean Astronomical Station (CAS). The BVRI data correspondent to the period 2013-2018 were obtained at CrAO.}
\label{tab:table05}
\begin{tabular}{ccccccccccc} 
\hline
\hline
HJD-2400000& B        & V       & R       & I       & J       & H       & K     & L     & M     \\
\hline
57313.551 & 10.40 & 9.44  & 8.54 & 7.95 & 7.35 & 6.69 & 6.00 & 5.14 & 4.74 \\
57324.290 & 10.77 & 9.75  & 8.79 & 8.11 & 7.47 & 6.77 & 6.02 & 5.06 & 4.80 \\
57391.301 & 10.30 & 9.38  & 8.49 & 7.92 & ...  & ...  & 5.99 & 5.12 & 4.98 \\
57399.372 & 10.27 & 9.34  & 8.44 & 7.91 & 7.36 & 6.67 & 6.03 & 5.12 & 4.99 \\
57426.268 & 10.53 & 9.53  & 8.64 & 8.03 & 7.36 & 6.65 & 6.01 & 5.05 & 4.78 \\
57623.524 & 10.16 & 9.33  & 8.42 & 7.86 & 7.27 & 6.57 & 5.94 & 5.05 & 5.16 \\
57635.600 & ...   & ...   & ...  & ...  & 7.25 & 6.59 & 5.96 & 5.11 & 4.91 \\
57638.500 & ...   & ...   & ...  & ...  & 7.32 & 6.66 & 6.02 & 5.13 & 4.86 \\
57641.504 & 10.23 & 9.42  & 8.46 & 7.88 & 7.31 & 6.62 & 5.98 & 5.12 & 4.89 \\
57648.474 & 10.35 & 9.48  & 8.51 & 7.95 & 7.37 & 6.69 & 6.02 & 5.12 & 4.92 \\
57680.500 & ...   & ...   & ...  & ...  & 7.26 & 6.56 & 5.87 & 4.99 & 4.81 \\
57704.600 & 10.10 & 9.23  & 8.38 & 7.83 & 7.22 & 6.54 & 5.92 & 5.08 & 4.99 \\
57730.327 & 10.08 & 9.20  & 8.39 & 7.85 & 7.23 & 6.60 & 5.93 & 5.06 & 4.87 \\
57745.456 & 10.08 & 9.24  & 8.40 & 7.84 & 7.30 & 6.72 & 6.08 & 5.23 & 5.03 \\
57746.448 & 10.06 & 9.21  & 8.42 & 7.86 & 7.31 & 6.69 & 6.07 & 5.21 & 5.07 \\
57774.399 & 10.02 & 9.21  & 8.36 & 7.82 & 7.26 & 6.67 & 6.05 & 5.18 & 4.98 \\
57777.256 & 10.02 & 9.19  & 8.35 & 7.78 & 7.22 & 6.60 & 5.99 & 5.13 & 5.04 \\
57785.257 & 10.04 & 9.22  & 8.37 & 7.83 & 7.28 & 6.66 & 6.04 & 5.20 & 5.10 \\
57786.265 & 10.10 & 9.25  & 8.41 & 7.86 & 7.31 & 6.68 & 6.06 & 5.20 & 5.06 \\
57795.232 & 10.17 & 9.33  & 8.44 & 7.89 & 7.38 & 6.72 & 6.02 & 5.19 & 5.00 \\
57803.187 & 10.46 & 9.53  & 8.59 & 8.01 & 7.42 & 6.73 & 6.04 & 5.14 & 5.05 \\
57812.300 & ...   & ...   & ...  & ...  & 7.44 & 6.76 & 6.07 & 5.14 & 5.06 \\
57817.264 & ...   & ...   & ...  & ...  & 7.34 & 6.69 & 6.04 & 5.12 & 4.96 \\
58004.501 & 10.16 & 9.30  & 8.43 & 7.90 & 7.25 & 6.62 & 5.96 & 5.09 & 4.95 \\
58008.546 & ...   & ...   & ...  & ...  & 7.26 & 6.62 & 5.95 & 5.03 & 4.77 \\
58026.556 & ...   & ...   & ...  & ...  & 7.23 & 6.58 & 5.92 & 5.00 & 4.78 \\
58038.401 & 10.49 & 9.55  & 8.64 & 8.07 & 7.35 & 6.66 & 5.96 & 5.00 & 4.79 \\
58066.348 & 10.55 & 9.61  & 8.66 & 8.10 & 7.34 & 6.64 & 5.90 & 4.91 & 4.69 \\
58096.335 & 10.38 & 9.45  & 8.55 & 8.00 & 7.29 & 6.61 & 5.89 & 4.87 & 4.60 \\
58100.369 & ...   & ...   & ...  & ...  & 7.35 & 6.66 & 5.91 & 4.92 & 4.64 \\
58114.373 & 10.60 & 9.64  & 8.69 & 8.13 & 7.36 & 6.65 & 5.90 & 4.84 & 4.54 \\
58120.191 & 10.66 & 9.67  & 8.71 & 8.12 & 7.31 & 6.63 & 5.87 & 4.87 & 4.63 \\
58125.323 & 11.07 & 10.00 & 8.96 & 8.33 & 7.53 & 6.75 & 5.98 & 4.93 & 4.75 \\
58126.193 & 10.94 & 9.88  & 8.88 & 8.25 & 7.46 & 6.73 & 5.96 & 4.91 & 4.67 \\
58143.030 & 12.03 & 10.82 & 9.64 & 8.86 & 7.84 & 6.94 & 6.06 & 4.93 & 4.57 \\
58144.238 & 11.95 & 10.72 & 9.56 & 8.79 & 7.81 & 6.94 & 6.03 & 4.94 & 4.64 \\
58151.174 & 10.88 & 9.83  & 8.83 & 8.21 & 7.42 & 6.70 & 5.90 & 4.88 & 4.61 \\
\hline
\end{tabular}
\end{table*}

\bsp	
\label{lastpage}
\end{document}